\documentclass[10pt,superscriptaddress,preprintnumbers, aps,pra,twocolumn, nofootinbib,floatfix]{revtex4-2}
\usepackage{amssymb,amsmath}
\usepackage{color}
\usepackage{hyperref}
\usepackage{listings}
\usepackage{physics}
\usepackage{amsmath}
\usepackage{graphicx}
\usepackage{bbold}
\usepackage{qcircuit}
\usepackage{bm}
\usepackage{amsfonts}
\usepackage{comment}
\usepackage{amsmath}
\usepackage{natbib}
\usepackage{multirow}
\usepackage{amsthm}
\usepackage[dvipsnames,usenames]{xcolor}
\usepackage{forest}
\usepackage{orcidlink}
\usepackage[normalem]{ulem}

\usepackage[normalem]{ulem}
\usepackage[dvipsnames,usenames]{xcolor}

\hypersetup{
    unicode=false,          
    pdftoolbar=true,        
    pdfmenubar=true,        
    pdffitwindow=false,     
    pdfstartview={FitH},    
    pdfauthor={},     
    colorlinks=true,       
    linkcolor=blue,          
    citecolor=blue,        
}

\graphicspath{{figures/}}

\begin{document}

\definecolor{dkgreen}{rgb}{0,0.6,0}
\definecolor{gray}{rgb}{0.5,0.5,0.5}
\definecolor{mauve}{rgb}{0.58,0,0.82}

\lstset{frame=tb,
  	language=Matlab,
  	aboveskip=3mm,
  	belowskip=3mm,
  	showstringspaces=false,
  	columns=flexible,
  	basicstyle={\small\ttfamily},
  	numbers=none,
  	numberstyle=\tiny\color{gray},
 	keywordstyle=\color{blue},
	commentstyle=\color{dkgreen},
  	stringstyle=\color{mauve},
  	breaklines=true,
  	breakatwhitespace=true
  	tabsize=3
}

\preprint{Preprint number IQuS@UW-21-043; LLNL-JRNL-853700}
    \title{Demonstration of a quantum-classical co-processing protocol for simulating nuclear reactions}
\newcommand{\iqusfil}{InQubator for Quantum Simulation (IQuS), Department of Physics,
University of Washington, Seattle, Washington 98195, USA}
\newcommand{\llnlafil}{Lawrence Livermore National Laboratory, P.O. Box 808, L-414, Livermore, California 94551, USA}
\newcommand{\utrentoafil}{Physics Department, University of Trento, Via Sommarive 14, I-38123 Trento, Italy}
\newcommand{\infnafil}{INFN-TIFPA Trento Institute of Fundamental Physics and Applications,  Trento, Italy}
\newcommand{\lbnlafil}{Applied Mathematics and Computational Research Division, Lawrence Berkeley National Laboratory, Berkeley, California 94720, USA}
\newcommand{\lbnlmatfil}{Materials Science Division, Lawrence Berkeley National Laboratory, Berkeley, California 94720, USA}
\newcommand{\qnlafil}{Quantum Nanoelectronics Laboratory, Department of Physics, University of California, Berkeley, California 94720, USA}

\author{F.~Turro \orcidlink{https://orcid.org/0000-0002-1107-2873}}
\thanks{These authors contributed equally. Email correspondence: fturro@uw.edu; trevor\_chistolini@berkeley.edu}
\affiliation{\iqusfil}
\affiliation{\utrentoafil}
\affiliation{\infnafil}

\author{T.~Chistolini\orcidlink{https://orcid.org/0000-0002-6461-2263}}
\thanks{These authors contributed equally. Email correspondence: fturro@uw.edu; trevor\_chistolini@berkeley.edu}
\affiliation{\qnlafil}

\author{A.~Hashim\orcidlink{https://orcid.org/0000-0002-8611-0125}}
\author{Y.~Kim\orcidlink{https://orcid.org/0000-0002-6332-1050}}
\author{W.~Livingston\orcidlink{https://orcid.org/0000-0001-8399-0975}} 
\affiliation{\qnlafil}
\affiliation{\lbnlafil}

\author{J.~M.~Kreikebaum\orcidlink{https://orcid.org/0000-0002-0253-4183}}
\affiliation{\qnlafil}
\affiliation{\lbnlmatfil}

\author{K.~A.~Wendt \orcidlink{https://orcid.org/0000-0002-3428-6479}}
\affiliation{\llnlafil}

\author{J.~L Dubois \orcidlink{https://orcid.org/0000-0003-3154-4273}}
\affiliation{\llnlafil}

\author{F.~Pederiva \orcidlink{https://orcid.org/0000-0002-7242-0042}}
\affiliation{\utrentoafil}
\affiliation{\infnafil}

\author{S.~Quaglioni \orcidlink{https://orcid.org/0000-0002-7512-605X}}
\affiliation{\llnlafil}
\date{\today}

\author{D.~I.~Santiago \orcidlink{https://orcid.org/0000-0001-8074-5130}}
\affiliation{\lbnlafil}

\author{I.~Siddiqi \orcidlink{https://orcid.org/0000-0003-2200-1090}}
\affiliation{\qnlafil}
\affiliation{\lbnlafil}
\affiliation{\lbnlmatfil}

\begin{abstract}
Quantum computers hold great promise for exact simulations of nuclear dynamical processes (e.g., scattering and reactions), which are paramount to the study of nuclear matter at the limit of stability and in the formation of chemical elements in stars. 
However, quantum simulations of the unitary (real) time dynamics of fermionic many-body systems require a currently prohibitive number of reliable and long-lived qubits. We propose a co-processing algorithm for the simulation of real-time dynamics in which the time evolution of the spatial coordinates is carried out on a classical processor, while the evolution of the spin degrees of freedom is carried out on quantum hardware. We demonstrate this hybrid scheme with the simulation of two neutrons scattering at the Lawrence Berkeley National Laboratory's Advanced Quantum Testbed. 
After implementing error mitigation strategies to improve the accuracy of the algorithm in addition to a combination of circuit compression techniques and tomography as methods to elucidate the onset of decoherence, our results validate the principle of the proposed co-processing scheme. 
A generalization of this present scheme will open the way for (real-time) path integral simulations of nuclear scattering.
\end{abstract}

\maketitle

\section{Introduction}
The simulation of the continuum properties of a general quantum many-body system remains one of the most challenging unsolved problems in theoretical physics.  Such problems encompass a broad range of phenomena from nuclear reactions and decays that drive stellar evolution~\cite{doi:10.1146/annurev-nucl-020620-063734,doi:10.1146/annurev-astro-081811-125543,RevModPhys_83_195}, to the electron-phonon scattering underlying superconductivity in materials~\cite{RevModPhys_89_015003,doi:10.1080/15567265.2021.1902441}.  These simulations are particularly difficult when the interaction between constituent bodies is nonperturbative, and the resources required to perform the necessary computations scale exponentially in the number of degrees of freedom.

The recent expansion of accessible quantum computing platforms~\cite{huang2020superconducting,bruzewicz2019trapped,henriet2020quantum} offers tools to explore new solutions to this extremely difficult problem~\cite{Jordan-Lee-Preskill-2012,liu2021towards}.  This is largely due to the capability of quantum computers to encode and process an exponentially growing Hilbert space as the number of qubits increases linearly.  
However, the age of robust, scalable, and fault-tolerant quantum computing is still on the horizon. Contemporary quantum platforms are hindered by having limited qubits and short executable circuit depths, placing a bound on the scale and complexity of applicable quantum algorithms. Until accessible platforms overcome these limitations, algorithms that are hybrid in nature~\cite{Peruzzo-2014,mcclean2016theory,google2020hartree,motta2020determining,mcardle2019variational}, using both classical and quantum computing resources, will likely offer the largest advantage over traditional computing.

In this paper, we explore a hybrid algorithm for computing the real-time dynamics of scattering fermionic many-body systems. We examine the particular case where the pairwise interaction Hamiltonian depends nonperturbatively on the spin state.  This case arises prominently in nuclear physics where central (spin-independent), spin-orbit, and even spin-tensor forces contribute with nearly equal strength~\cite{RevModPhys.81.1773}. 
Furthermore, the nuclear case depends similarly nonperturbatively on the isospin, a synthetic degree of freedom that encodes whether a nucleon is a proton or a neutron.  This provides another source of exponential growth in the degrees of freedom required to describe the nuclear many-body system, making relevant simulations increasingly challenging.  Indeed, the many-body propagator that generates the dynamical evolution of the many-body wave function is a matrix that must encompass all possible spin and isospin states of the system, the dimension of which grows exponentially with the  number of nucleons.  The explosion of the possible states is the primary limiting factor in applying "classical" imaginary time evolution algorithms, such as the Green's Function Monte Carlo, to structure calculations of nuclei made of more than about 12 nucleons~\cite{RevModPhys.87.1067}.

In general, the propagator for both real and imaginary time evolution can be approximately factorized by means of the Trotter decomposition in order to isolate specific parts of the Hamiltonian~\cite{Trotter-1959,suzuki-1993}.  Often, this is done to separate the impact of the kinetic energy and the interaction Hamiltonian on the propagator, with each part being treatable exactly in conjugate spaces. In contrast, we split the Hamiltonian into terms that are independent of particle spin and act only on the spatial degrees of freedom, and those that also depend explicitly on the particle spins. 
We propose a hybrid scheme where the propagator containing spin dependent terms is enacted using quantum hardware, while the purely spatial propagation is evaluated on a classical processor.
We stress that the propagator acting on the spin degrees of freedom is a function of the particles' relative coordinates and hence changes along their trajectories in the scattering process.
In this context, and as a proof of principle demonstration, we use a relatively simple model for the spatial propagation and focus on the real time spin evolution. We aim to test the feasibility of
computing the outgoing occupation probabilities in the asymptotic region, or in other words the probability of scattering from the initial to the final state of the system, through the repeated application of a nontrivial spin propagator that changes with the spatial evolution. The time, and hence number of time steps, required to reach the asymptotic scattering region are dictated by the Hamiltonian of the system under consideration. Furthermore, we exclusively focus on the Hamiltonian propagation and not on the problem of preparing the initial state of the system. 
Different techniques designed to compute the ground state of a generic Hamiltonian~\cite{Turro2022,QITE_JL, QITE_Jouzdani,choi2021rodeo,perez2022quantum,di2021improving,hlatshwayo2022simulating,stetcu2022variational,dumitrescu2018cloud} can be preliminarily employed to achieve a desired physical initial state. 

Through a user project at the Lawrence Berkeley National Laboratory, Advanced Quantum Testbed (AQT)~\cite{Berkeley_AQT}, we carried out an initial demonstration of this hybrid co-processing scheme for the scattering of two neutrons, where we factorize the spin-dependent and spin-independent (or spatial) degrees of freedom for the short-time evolution (referred to as the adiabatic approximation).

Until now, quantum algorithms for the simulation of nuclear dynamics have been explored in the context of calculations of linear-response functions~\cite{roggero2019dynamic,roggero2020quantum,baroni2022nuclear,inelastic_scattering} to electroweak probes or another external time-dependent field, radiative processes~\cite{bedaque2022radiative}, and  Green’s functions~\cite{Anthony_scattering}.
We thus apply quantum computing towards simulating nuclear reactions that employs a real-time approach.
Indeed, in this work, we aim to establish the groundwork for simulating a nuclear scattering experiment with a quantum real-time evolution algorithm.
We demonstrate the validity of the proposed quantum-classical co-processing scheme, especially after implementing multiple error mitigation strategies that also diagnose the effects of decoherence.

This paper is organized as follows: Section~\ref{sec:theory} describes the theoretical background of our simulations and outlines the proposed and experimentally validated algorithm. Section~\ref{sec:methods} describes the methods used to implement the spin evolution on the quantum processor, focusing on the error mitigation strategies necessary for an accurate simulation. Section~\ref{sec:results} reports the obtained experimental data and results that demonstrate the success of this scheme. Finally, Section~\ref{sec:conclusions} reports the conclusions.

\section{Theoretical background of nuclear scattering \label{sec:theory} }

\begin{figure}
    \centering
    \includegraphics[scale=0.3]{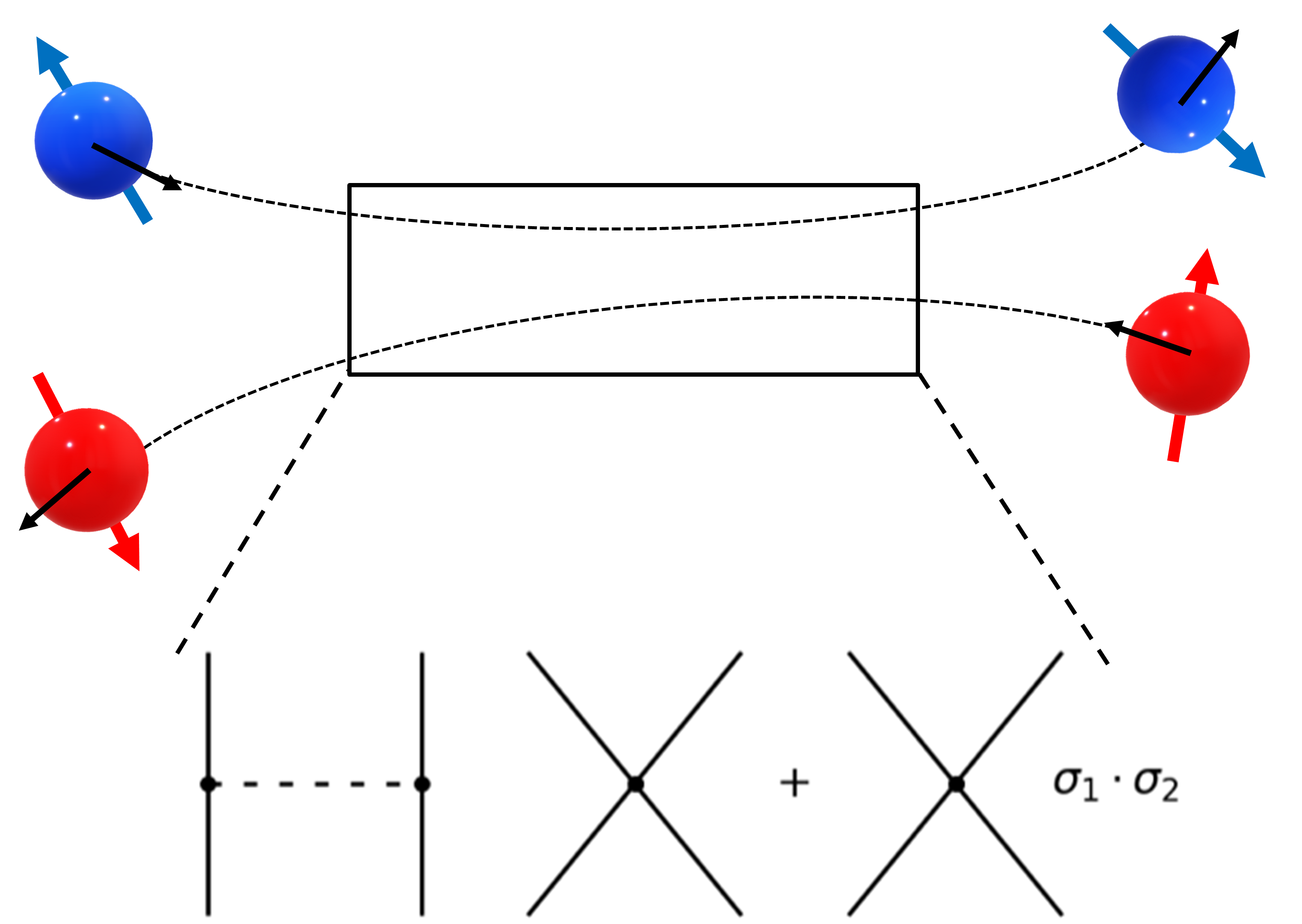}
 \caption{Pictorial representation of the scattering between two nucleons. The two vectors denote the associated spin and direction. Lower graphs depict the Feynman diagrams of the LO $\chi$-EFT. }
    \label{fig:Feynman}
\end{figure}

The simplest yet realistic model for the nucleon-nucleon interaction can be obtained from the leading order (LO) of a chiral effective field theory ($\chi$-EFT). This interaction is based on the Feynman diagrams in the lower part of Fig.~\ref{fig:Feynman}, while a conceptual illustration of the two-neutron scattering event is portrayed in the upper part. From left to right, the first Feynman diagram describes the one-pion exchange process and the second and third diagrams represent contact interactions. The corresponding Hamiltonian can be written as a sum of three terms: 
\begin{equation}
H\,=\,T\,+\,V_{\rm SI}\,+\,V_{\rm SD}\,.\label{eq:nuclear_Ham}
\end{equation}
The first term, $T$, represents the kinetic energy, while $V_{\rm SI}$ describes the spin-independent portion of the nucleon-nucleon potential.
The remaining spin-dependent component of the interaction (acting on the spin degrees of freedom) is denoted as $V_{\rm SD}$, where more explicitly it can be written as
\begin{equation}
\begin{array}{rcl}
        V_{SD}(\vec{r}) &=& 
            V_{s}(r) \; \bm{\sigma}_1 \cdot \bm{\sigma}_2 \,\\ \\
            && +V_{t}(r)
            \left(
            \bm{\sigma}_1 \cdot \hat{r} \; \bm{\sigma}_2 \cdot \hat{r} - 
            \displaystyle\frac{ 
                \bm{\sigma}_1 \cdot \bm{\sigma}_2}
                {3}
            \right),
        \end{array}
        \end{equation}
where $\vec{r}=\vec{r}_2-\vec{r}_1$ is the nucleon-nucleon distance.
Accordingly, the time evolution operator can be approximated as
\begin{equation}
e^{-i\Delta t H}\,\simeq \,e^{-i \Delta t \left(V_{\rm SI}+T\right)} \,e^{-i \Delta t \,V_{\rm SD}} + O(\Delta t^2)
\,,\label{eq:coprocessingevolution}
\end{equation}
where we have used the Trotter decomposition to split the propagator in the limit of an infinitesimal time interval, $\Delta t\xrightarrow[]{}0$. 
At each time step, the $e^{-i \Delta t \left(V_{\rm SI}+T\right)}$ propagator evolves the spatial degrees of freedom, while $e^{-i \Delta t \,V_{\rm SD}}$ evolves the spin degrees of freedom.
This type of Trotter decomposition was first introduced in Ref.~\cite{Holland-2020}, which modeled the output of 
a superconducting qudit processor implementing the (simpler) time evolution of two neutrons' spins frozen in space by means of custom quantum gates.   

We develop a hybrid quantum-classical algorithm for the simulation of the real time evolution of two neutrons, in which the evolution of the spatial coordinates is carried out on a classical processor, while the evolution of the spin degrees of freedom is conducted on a quantum processor. Specifically, we implement the following scheme:
\begin{enumerate}
    \item Map the spin states of the two neutrons onto the computational states of a two qubit processor. In this application, we used two qubits of an eight qubit superconducting quantum processor at the AQT, where further details can be found in Ref. \cite{RC_Akel} and in Appendix \ref{app:hardware}. 
    We use the $|00\rangle$, $|01\rangle$, $|10\rangle$, and $|11\rangle$ states to represent, respectively, the coupled two-spin states $| S=1, S_z=-1 \rangle = |\downarrow\downarrow\rangle$, $| S=1, S_z=0 \rangle =\frac{1}{\sqrt{2}}\left(|\uparrow\downarrow\rangle+|\downarrow\uparrow\rangle\right)$, $| S=1, S_z=1 \rangle = |\uparrow\uparrow\rangle$, and $| S=0, S_z=0 \rangle = \frac{1}{\sqrt{2}}\left(|\uparrow\downarrow\rangle-|\downarrow\uparrow\rangle\right)$.
    \item At each time step $\Delta\,t$:
    \begin{enumerate}
        \item Evolve the spatial coordinates (using a standard algorithm) on a classical device.
        \item Construct the short-time spin propagator that evolves the spin states from time $t$ to time $t+\Delta\,t$ at the updated relative coordinate, $\vec{r}(t)$.
    \item Compile this propagator as a quantum circuit and execute it on the quantum device to propagate the spin degrees of freedom. 
    \item Repeat from step (a) for the evolution's duration.
    \end{enumerate}    
\end{enumerate}
At leading order of chiral EFT, the spatial propagator does not depend explicitly on the spin states (see Appendix~\ref{app:spin_proof} for a proof). Therefore, we can further simplify the simulation protocol by independently computing the spatial evolution before the quantum simulation. 

Here, the spatial trajectory is  obtained by solving the Newton equation. This simplification was not meant to attain the most realistic results but rather to test the viability of the method, and in particular the feedback between the classical and quantum sides of the algorithm. All the same, from a quantum point of view, the obtained classical trajectory describes the most probable path of the two neutrons guaranteed by the saddle-point approximation~\cite{Feynman_saddlepoint}. 

\section{Quantum Circuit Implementation for Co-Processing Algorithm}\label{sec:methods}

\begin{figure*}[t]
\includegraphics[scale=0.38]{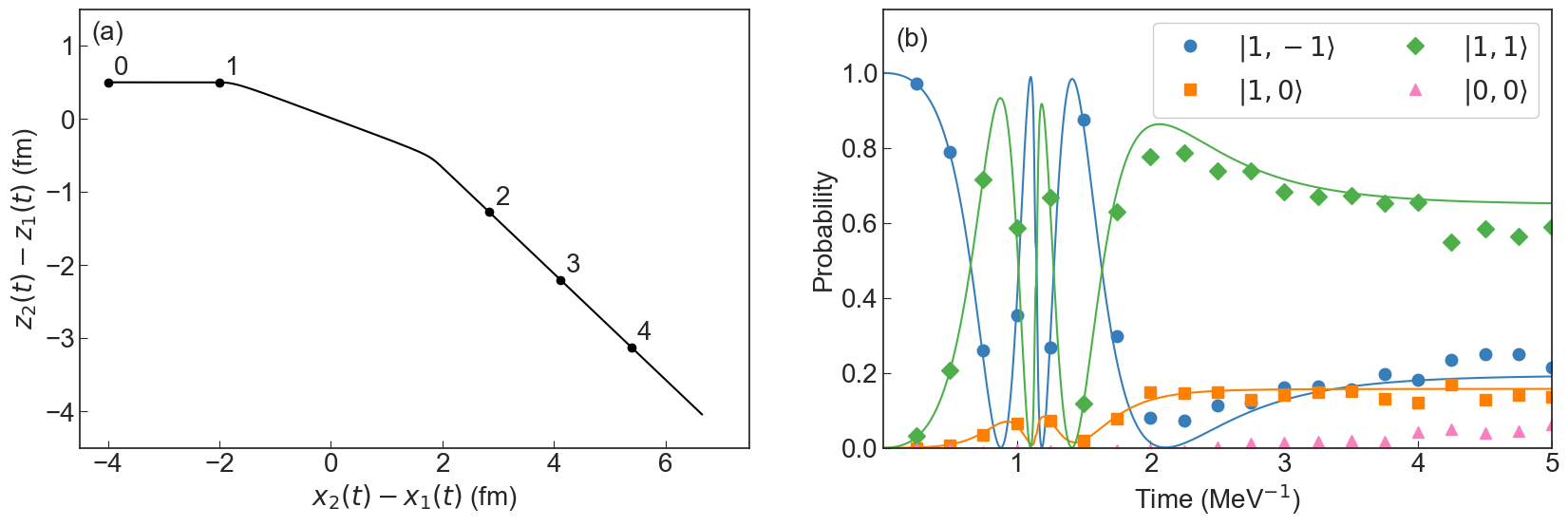}
\caption{Trajectory in the relative frame in the $x$-$z$ plane. The enumerated points each correspond to the position at time $t\,=\,0,\,1,\,2,\,3,\,4$ MeV$^{-1}$. In the right panel, we present the best results of the spin evolution. The points indicate the experimental results, while the lines represent the exact Trotter decomposition evolution. Readout calibration, randomized compiling, and state purification methods were used to mitigate noise.} \label{fig:Coprocessing_Spatial_time}
\end{figure*}

The spatial trajectory, $\vec{r}(t)$, of the two neutrons is computed by solving the Newton equation for spatial dynamics employing the Verlet integration~\cite{verlet1967computer,swope1982computer}. For this evolution, we employed a total of $N = 1000$ time steps of size $\Delta  t=0.005$ MeV$^{-1}$.  Figure~\ref{fig:Coprocessing_Spatial_time}(a) shows the results in the $x$-$z$ plane, where $x=x_1-x_2$ and $z=z_1-z_2$ are the components of the relative coordinate as a function of time.
Having the relative positions in time, we then computed the short time spin propagators, $e^{-i\,\Delta t\,V_{\rm SD}(\vec{r}(t))}$.
\begin{figure}[th]
    \centering
            $$ \Qcircuit @C=1em @R=.7em {
     & \multigate{1}{U(t_0)} & \qw& \multigate{1}{U(t_1)} & \qw& \multigate{1}{U(t_2)} & \qw& \multigate{1}{U(t_3)} & \qw & \multigate{1}{U(t_4)} & \qw\\
     & \ghost{U(t_0)} & \qw & \ghost{U(t_1)} & \qw  & \ghost{U(t_2)} & \qw  & \ghost{U(t_3)} & \qw  & \ghost{U(t_4)} & \qw
} $$ \\
    $$ \Qcircuit @C=1em @R=.7em { 
    &\multigate{1}{U_{reini}(T_{reini})} & \qw &
    \multigate{1}{U(t_3)} & \qw & \multigate{1}{U(t_4)} & \qw\\
     &  \ghost{U_{reini}(T_{reini})} & \qw& \ghost{U(t_3)} & \qw  & \ghost{U(t_4)} & \qw
} $$ 
\caption{Schemes for the quantum circuits simulating the spin evolution up to time step $j=4$. Top: the implemented quantum circuit using a sequence of propagators. Bottom: the quantum circuit employed in the reinitialization scheme. The $U_{\rm reini}$ operation represents the reinitialization gate.}
\label{fig:qc_reinitializinggates}
\end{figure}
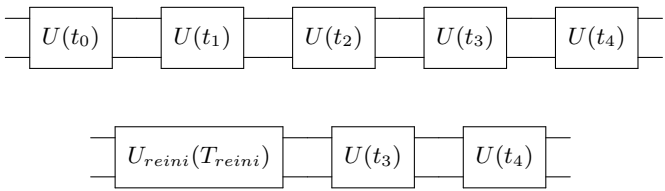 
\begin{figure*}
\centering
\includegraphics[scale=0.38]{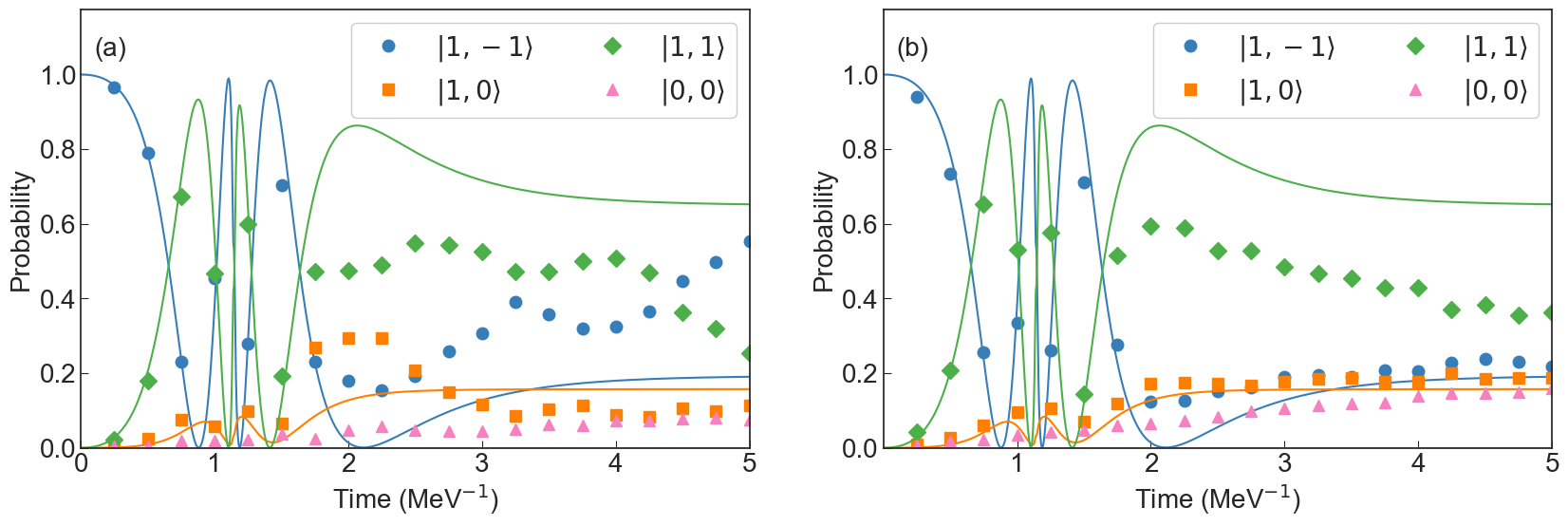}
\caption{Initial results of the spin evolution. The points indicate the experimental results, while the lines represent the exact Trotter decomposition evolution. The left panel (a) shows results where no error mitigation methods are applied.  The right panel(b) shows results after applying RCAL and RC to mitigate noise.}
\label{fig:initial_evolution1}
\end{figure*}
For the spin evolution, we reduced the total number of time steps to $N_s=20$, from $T_0=0$ to $T_{N_s}=19$, by computing the short-time spin propagators along the spatial trajectory for 50 consecutive time intervals ($\Delta T = 50 \Delta t$) according to 
\begin{equation}
 U_{\rm coarse}(T_j)=  {\mathcal T}\exp\left[-i \int_{T_j}^{T_{j+1}} V_{\rm SD}(\vec{r}(t))dt\right]  \label{eq:Coprocessin_long_time_propagator},
\end{equation}
where $T_j= 50\,j \Delta t$.
Each circuit was designed to simulate the $j$th propagator $U_{\rm coarse}(T_j)$.  

The spin probability as a function of time  was obtained by concatenating the quantum circuit implementing the sequence of time-ordered operators $U_j$, with $j\,<\,i$, and executing it on the quantum processor. A scheme of the implemented quantum circuit is shown in the upper half of Fig.~\ref{fig:qc_reinitializinggates}. Each $U_j$ operator is compiled with digital gates, specifically, $R_x$, $R_z$, and CNOT, using the \textit{decompose} function of the \textit{Qiskit} open-source software development kit~\cite{qiskit}. For each $U_j$ operator, the quantum circuit is composed of three CNOT and eight $U_3$ gates. Thus, circuits simulating the spin evolution at longer times increase linearly in depth.

Our most accurate spin evolution using strategies that maintain the original circuit's depth is shown in Fig.~\ref{fig:Coprocessing_Spatial_time}(b). We observe that our results reach the exact asymptotic limit (after t=$4$ MeV$^{-1}$) where the spin state probabilities become constant because the two neutrons are far apart with a very weak interaction. Therefore, this demonstrates the feasibility of simulating a real-time scattering process using a quantum computer.
Initially though, without applying error mitigation methods, experimental results diverged from ideal results as the simulation progressed in time and depth, as illustrated in Fig.~\ref{fig:initial_evolution1}(a). 
To address this, error mitigation strategies were implemented to improve the accuracy of the algorithm and validate the principle of this co-processing scheme.

For the remainder of this section, we review the error mitigation strategies applied toward this simulation of a nuclear scattering event, and we also explore methods employing either circuit compression techniques or tomography as diagnostic tools to evaluate current limitations.

\subsection{Error mitigation methods}\label{subsec:methods_scalable}

There can be a host of errors diminishing the algorithm's accuracy, such as arising from imperfect state preparation and measurement (SPAM), coherent errors, and stochastic errors. To mitigate against readout errors, we applied the readout calibration (RCAL) protocol from Quantum Benchmark's True-Q\texttrademark \ package \cite{True_Q}. 
Randomized compiling (RC) was applied to translate coherent error to stochastic error \cite{RC_original,RC_Akel}. 
Once the noise could be accurately fit by an approximate depolarizing error model, a simple form of state purification was applied to renormalize Pauli expectation values, using only the measured two-qubit process fidelity instead of more expensive methods employing state tomography \cite{QITE_JL}. These methods were all utilized in order to improve the accuracy of the algorithm, and further details can be found in Appendix \ref{app:Err_Mitigation}.

\subsection{Reinitialization methods \label{sec:Reinitialization}}

In addition to these strategies, we also explored using circuit compression and tomography as methods to reduce the algorithm's circuit depth in attempt to achieve greater accuracy or to experimentally simulate finer dynamics. While these methods are not scalable, we explore them to investigate limiting factors in the current experiment in addition to testing new procedures.
The first technique was circuit compression, relying upon Cartan's KAK decomposition to reduce all circuits of the algorithm to include at most three $CNOT$ gates~\cite{Cartan}. With all circuits consequently having constant depth as opposed to increasing linearly with time, we were able to explore the algorithm's effectiveness beyond the limitations imposed by decoherence at longer circuit depths.

In addition to circuit compression, we developed and explored a reinitialization procedure employing state tomography~\cite{Nielsen_Chuang} after a chosen number of time steps, $T_{\rm reini}$, to overcome some effects of noise.  
Using tomography to measure the spin state at time $T_{\rm reini}$, we implemented a new quantum circuit composed of a state preparation circuit followed by circuits that further evolve the state in time. Iterating this method, we can study longer time dynamics with reasonable accuracy, as the circuit depth is periodically reduced instead of consistently growing in time. This reinitialization procedure is similar in spirit to the `restarting' procedure proposed in Ref.~\cite{otten2019noiseresilient}, where the state after one step of dynamical evolution is approximated by optimizing a variational circuit before performing the next time step. However, here, we employ a different state tomography algorithm, in which the (mixed) qubit states are approximated by pure ones. The description and details of the employed state tomography process and the construction of the reinitialization circuit, which in our present application is a two-qubit gate, can be found in Appendix~\ref{sec:Tomography}.

\section{Data and Results\label{sec:results}}
While the spatial evolution of the algorithm was performed using a classical computer, the spin evolution was conducted on a quantum processor. As described above, initial simulations of the algorithm yielded results that diverged away from the theoretical evolution quickly, shown most clearly in Fig.~\ref{fig:initial_evolution1}(a). 
\begin{figure}
\centering
\includegraphics[scale=0.38]{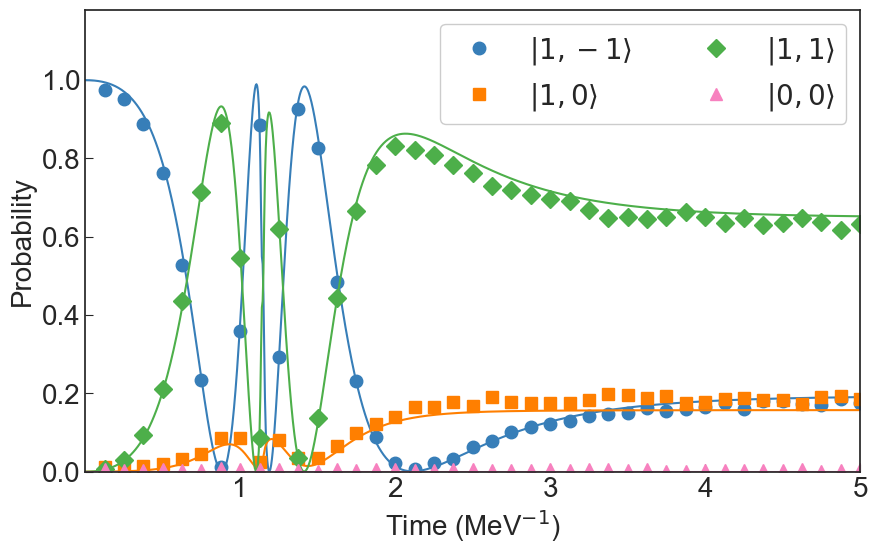}
\caption{Results after implementation of circuit compression, RCAL, RC, and state purification. 40 time steps are used to resolve more detailed dynamics.}
\label{fig:copression_plot}
\end{figure}
In response to these preliminary results, we began employing a series of error mitigation strategies, as described in Sec.~\ref{subsec:methods_scalable}. To correct for readout errors, the RCAL protocol from Quantum Benchmark's True-Q\texttrademark ~\cite{True_Q} package yielded concrete yet minor improvements to the algorithm performance. 
RC was implemented to translate coherent error to incoherent error, demonstrating further improvements. All the same, as seen in Fig.~\ref{fig:initial_evolution1}(b), while RCAL and RC both lead to improved accuracy, the state probabilities do not enter the asymptotic regime but continue converging towards each other.
Thus, state purification was used, which when combined with these other two error mitigation protocols, acted as a reliable method to extend the algorithm out to the desired time duration with $20$ time steps. 

While improvements were made by implementing these error mitigation protocols, we continued to explore methods to attain further accuracy in the algorithm's performance. Although not scalable to higher numbers of qubits, circuit compression had a dramatic impact. It enabled us to either probe the spin evolution over longer time periods with the same number of steps, or to use a finer discretization of time steps over the same time length to evaluate finer dynamics, both without running into limits of decoherence. Fig.~\ref{fig:copression_plot} displays results when circuit compression is added to the above error mitigation strategies, and $40$ time steps are used. This technique also serves to convey the increased viability of such a simulation protocol when further improvements are eventually made to qubit coherence.

To quantify the performance of the algorithm and more clearly compare the effectiveness of the error mitigation strategies, we use the total variation distance (TVD)~\cite{TVD_photons}. The TVD, in Eq.~\eqref{TVD_eqn} written as $D(P,Q)$, quantifies the absolute difference between two probability distributions $P$ and $Q$, which here are the experimental and ideal results. Mathematically, 
\begin{equation}
    D(P,Q) = \frac{1}{2}\sum_i \abs{P_i - Q_i}\,.
\label{TVD_eqn}
\end{equation}

\begin{figure}
    \centering
    \includegraphics[scale=0.38]{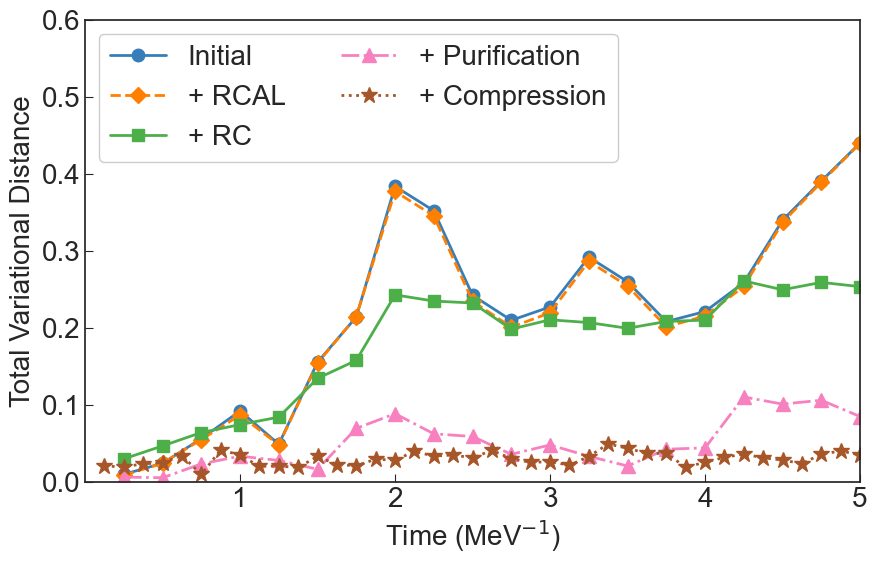}
    \caption{Total variational distance as successive error mitigation strategies are implemented.}
    \label{fig:TVD_summary}
\end{figure}

We plot the TVD of the results from applying a series of error mitigation strategies in Fig.~\ref{fig:TVD_summary}. As discussed, while initial trials of the spin evolution algorithm with no error mitigation methods diverged away from ideal results relatively quickly, incorporating this set of strategies greatly improved the accuracy and enabled us to arrive at the asymptotic regime. This improvement in accuracy is especially seen at later time steps, as the circuits become deeper without use of circuit compression. As expected, incorporating circuit compression yielded the most accurate results.

\begin{figure}[ht]
    \centering
    \includegraphics[scale=0.38]{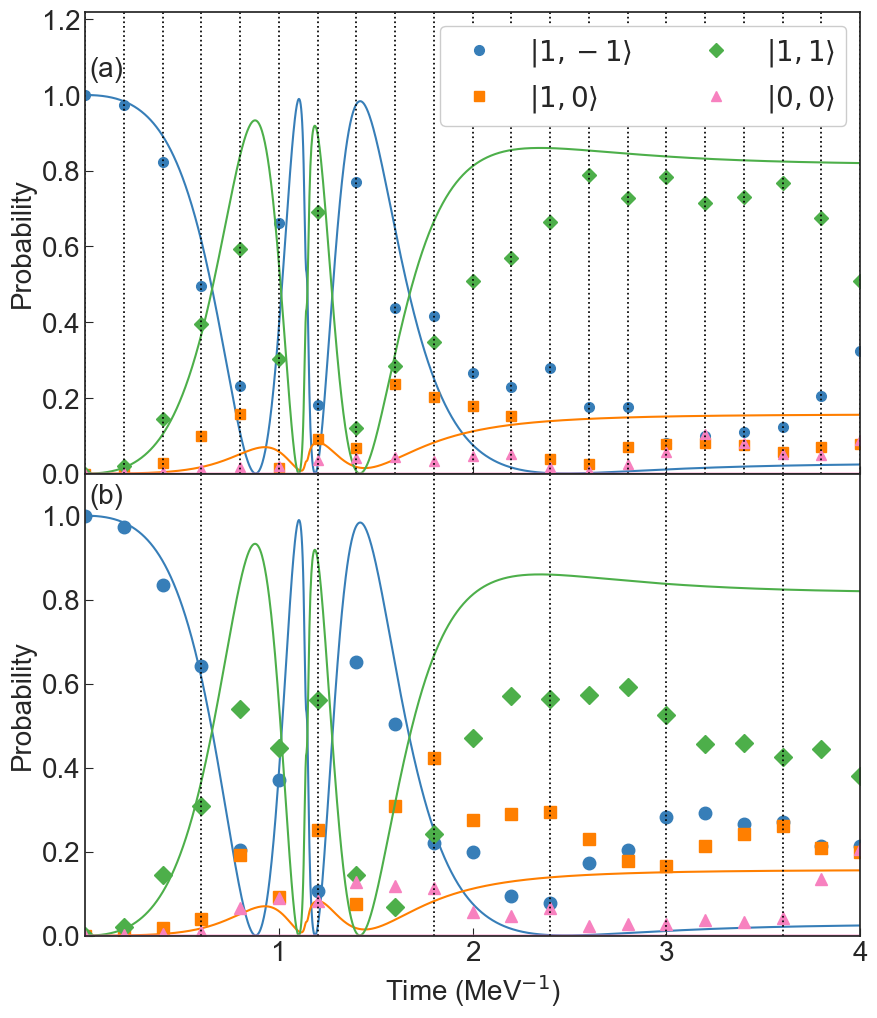}
    \caption{Co-processing results with the reinitialization method. Vertical dashed lines indicate the steps when there is a reinitializing process. The upper panel (a) shows results with $T_{\rm reini}=1,2,3,\ldots\times \Delta{}T$. The lower panel (b) shows results with  $T_{\rm reini}=3,6,9,\ldots\times \Delta{}T$.}
    \label{fig:prob_tomography_Tini1}
\end{figure}

Finally, we now discuss the quantum simulation results when implementing the reinitialization procedure discussed in Sec.~\ref{sec:Reinitialization} and schematically represented in Fig.~\ref{fig:qc_reinitializinggates}.
Using this method, we obtained the results presented in Fig.~\ref{fig:prob_tomography_Tini1}, where the reinitialization procedure is applied at each time step in panel (a), or it is applied every third time step in panel (b). Points represent the obtained spin state occupation data, solid lines display the solution of the exact Trotter decomposition evolution, and vertical dashed lines indicate when the reinitialization circuit was applied. No other error mitigation techniques were implemented.

\begin{figure}[ht]
    \centering
    \includegraphics[scale=0.38] {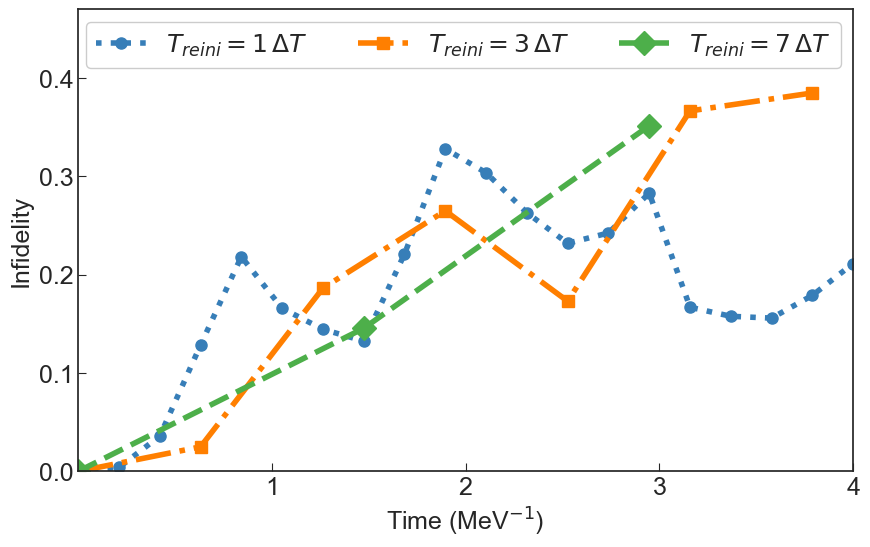}
    \caption{The infidelity between the obtained results using partial tomography and the exact Trotter decomposition states for different choices of $T_{\rm reini}$.}
    \label{fig:coprocessing_AQT_fidelity}
\end{figure}

Figure~\ref{fig:coprocessing_AQT_fidelity} summarizes the fidelities between the state computed from the exact Trotter-decomposed evolution and the experimental quantum simulation obtained via state tomography at $T_{\rm reini}=\,1,\,3,\,7 \times \Delta T$. This demonstrates that at short times, the co-processing algorithm is effective in reproducing the exact evolution, and different choices of $T_{\rm reini}$ perform similarly as the limiting factor is likely not circuit depth (as also seen in Fig. \ref{fig:TVD_summary}). It is only at longer times that the improved accuracy of $T_{\rm reini} = 1$ over the other choices is apparent, thus demonstrating the potential use of such a procedure that is more helpful as the times between reinitialization become shorter, as expected.
However, we acknowledge that the cost of performing state tomography scales exponentially with system size, and its applicability is therefore limited to small systems. Here, we used it as a diagnostic tool to investigate the performance of the present quantum-classical co-processing scheme.

\section{Conclusion\label{sec:conclusions}}

We present a study of simulating a nuclear scattering process utilizing a quantum processor, serving as a proof of concept of the technology's application toward nuclear processes. 
The reported co-processing protocol is a hybrid method where certain degrees of freedom are simulated on a quantum processor, while the remaining degrees of freedom are simulated using a classical computer. Specifically, we simulated the spin evolution of two-neutron scattering on the AQT quantum processor and the spatial evolution through a classical method that solves the Newton equation. 
Even with the simplicity of the demonstrative system studied, considering the results after application of error mitigation techniques that reduce the effect of noise, our results suggest that the present co-processing protocol may provide a promising pathway for simulating quantum-scattering experiments with a quantum computer.
Leveraging future quantum platforms with longer coherence times and higher gate fidelities, where it would be possible to avoid using nonscalable error mitigation methods, the direct real-time simulation of nuclear reaction experiments will be possible, enabling the robust computation of reaction properties, for example, cross sections, reaction rates, and branching ratios, of nuclei that are too short-lived to study in a laboratory.

From our initial results, we observed that various noise sources can pose a significant challenge to extending algorithms out to higher circuit depths. In response, we implemented various error mitigation techniques in addition to circuit compression and the described reinitialization method that both reduce the algorithm's depth, in order to diagnose the validity of our co-processing scheme. Ultimately, we find that the noise encountered could be sufficiently reduced by the described methods. In particular, once the noise could be captured by a depolarizing model after using RC, state purification yielded results demonstrating the spin state evolution reached the asymptotic regime.

A possible generalization of this co-processing scheme would study the full quantum simulation of scattering experiments. One could sample from the spatial distribution paths and use a quantum processor to evaluate the relevant spin probability or state. Summing all contributions from the paths, one could reach the quantum transition probability according to the path integral theory. However, this approach must overcome outstanding challenges in many-body simulations; some main ones include the difficulty of sampling from the real-time path integral and the treatment of the Fermion sign problem~\cite{PhysRevC.101.064001,Troyer_fermion_sign}. 
Moreover, although we neglect these problems here when simulating two-body dynamics, we must consider that the global phase of the final spin state for each path becomes a relative phase in the path summation. Therefore, it must be measured and saved to some register. Nevertheless, further study to test if this co-processing scheme can be generalized would be desirable.

\section*{Acknowledgements}

We would like to thank Alessandro Roggero, Piero Luchi, Valentina Amitrano, Marc Illa Subina, Anthony Ciavarella, Hersh Singh and Martin Savage for useful discussions and support.
This work was supported in part by the U.S. Department of Energy, Office of Science, Office of Nuclear Physics, InQubator for Quantum Simulation (IQuS) \cite{Iqus_website} under Award Number DOE (NP) Award DE-SC0020970 via the program on Quantum Horizons: QIS Research and Innovation for Nuclear Science.

T.C. acknowledges support by the National Science Foundation Graduate Research Fellowship Program under Grant No. DGE 1752814 and DGE 2146752.
This work was prepared in part by LLNL under Contract No.\ DE-AC52-07NA27344 with support from the Laboratory Directed Research and Development Grant No.\ 19-DR-005, and it was funded in part by the U.S. Department of Energy, Office of Science, Office of Nuclear Physics (under Work Proposal No.\ SCW1730), the Office of Advanced Scientific Computing Research Quantum Testbed Program (under Work Proposal No.\ FP00008338), and the Department of Energy National Nuclear Security Administration, Advanced Simulation and Computing Program. 

\appendix

\section{USING STATE TOMOGRAPHY AND THE REINITIALIZATION ALGORITHM \label{sec:Tomography} }
This work is based on Appendix F of Ref.~\cite{Turro2022}. We start by recapping it, and then we generalize it to the case of $n$ qubits. At the end of this section, we describe how to build the reinitializing operator from the results of state tomography.

\subsection{Tomography for 1 qubit}
Our implemented state tomography process for a single qubit is based on the following steps:
\begin{enumerate}
\item Measure the probabilities of the bare circuit. We thereby obtain $P_0$ and $P_1$.
\item Measure the probabilities of the bare circuit with a final $R_y(-\frac{\pi}{2})$ rotation.
\item Measure the probabilities of the bare circuit with a final $R_x(-\frac{\pi}{2})$ rotation.
\end{enumerate}
One can prove that the relative phase of a pure state between the two states is given by:
\begin{equation}
  \phi\,=\,\arctan\left(\frac{-P^x_0+\frac{1}{2}(P_0+P_1)}{P^y_0-\frac{1}{2}(P_0+P_1)}\right)\,,
\end{equation}
where $P^x_0$ and $P^y_0$ indicate the probability of measuring the $\ket{0}$ state after applying the $R_x$ and $R_y$ gates, respectively. $P_0$ and $P_1$ are the bare probabilities, obtained in Step 1.

\subsubsection{Generalization for n qubits}

For the case of $n$ qubits, we implement $2\,n+1$ quantum circuits to evaluate the state. The general state is given by $\left(P_0,\,P_1\,e^{i\phi_1},\,\ldots,\,P_{2^n-1}\,e^{i \phi_{2^n-1}} \right)^T$. The generalization of the presented state tomography process is obtained by following the below steps:
\begin{enumerate}
    \item Measure the bare probabilities $P_0,\dots,P_{2^n-1}$ in the computational basis, $\ket{0},\,\ket{1},\,\ldots,\,\ket{2^n-1}$.
    \item For $i=0,\,1,\,\ldots,\,n-1$:
    \begin{itemize}
        \item Measure the probabilities after applying $R_x$ and $R_y$ gates to qubit $n-1-i$, and applying the Identity to all others.
    \end{itemize}
\end{enumerate}
One obtains the following expression when we employ the rotation on the $n-1$ qubit, $R^0_{x(y)}=R_{x(y)}(-\frac{\pi}{2})\otimes\mathbb{1}\otimes \ldots\otimes \mathbb{1}$: 
\begin{equation}
 \phi_{2^{n-1}}-\phi_{0}=\,\arctan\left(\frac{-P^x_0+\frac{1}{2}(P_0+P_{2^{n-1}})}{P^y_0-\frac{1}{2}(P_0+P_{2^{n-1}})}\right)\,,
\end{equation}
where $P^x_0$ and $P^y_0$ indicate the probability of measuring the $\ket{0}$ state after applying the $R_x$ and $R_y$ gates, respectively. $P_0$ and $P_{2^{n-1}}$ are the bare probabilities, obtained in Step 1. We can set $\phi_0=0$ because it represents the global phase.

The second step of rotations, given by $R_{x(y)}^{1}=\mathbb{1} \otimes R_{x(y)} \otimes \ldots \otimes \mathbb{1} $,  yields the following expression:
\begin{equation}
\begin{split}
    \phi_{i+2^{n-2}}-\phi_{i}=&\,\arctan\left(\frac{-P^x_i+\frac{1}{2}(P_i+P_{i+2^{n-2}})}{P^y_i-\frac{1}{2}(P_i+P_{i+2^{n-2}})}\right) \\
    &i=0,\,2^{n-1}\,,
\end{split}
\end{equation}
where, as before, $P^x_j$ and $P^y_j$ are the probabilities of measuring the state $\ket{j}$ after applying the $R_x$ and $R_y$ gates, respectively. $P_k$ represents the bare probability of state $\ket{k}$ obtained in Step 1. The values of $\phi_i$ with $i=0,\,2^{n-1}$ are already computed in the previous step.

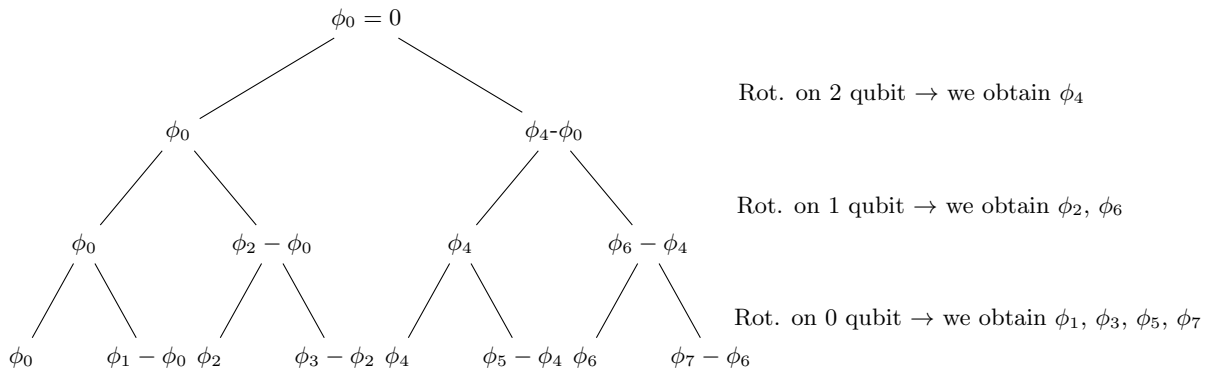
\begin{figure*}[t]
\begin{tikzpicture}[level/.style={sibling distance=50mm/#1}]
\node[] (z){$\phi_0=0$}
  child {node[] (a) {$\phi_0$}
    child {node[] (b) {$\phi_0$}
      child {node[] (d) {$\phi_0$}} 
      child {node[] (e) {$\phi_1-\phi_0$}}
    }
    child {node[] (g) {$\phi_2-\phi_0$}
      child {node[] (i) {$\phi_2$}}
      child {node[] (h) {$\phi_3-\phi_2$}}
    }
  }
  child {node[] (a) {$\phi_4$-$\phi_0$}
    child {node[] (b) {$\phi_4$}
      child {node[] (d) {$\phi_4$}} 
      child {node[] (e) {$\phi_5-\phi_4$}}
    }
    child {node[] (g) {$\phi_6-\phi_4$}
      child {node[] (i) {$\phi_6$}}
      child {node[] (h) {$\phi_7-\phi_6$}
    }
    }
  };
 \node[] at (7.25,-1) (l0) {Rot. on 2 qubit $\rightarrow$ we obtain $\phi_4$};
\node[] at (7.5,-2.5) (l1) {Rot. on 1 qubit $\rightarrow$ we obtain $\phi_2,\,\phi_6$ };
\node[] at (8,-4) (l2){Rot. on 0 qubit $\rightarrow$ we obtain $\phi_1,\,\phi_3,\,\phi_5,\,\phi_7$ };
\end{tikzpicture}
\caption{This decision tree shows how we link the relative phases with the rotations in the case of three qubits. Here, $\phi_0$ represents the global phase, so we can set $\phi_0=0$.} 
\label{fig:pyramid_relative_phases}
\end{figure*}

The last step of rotations, given by $R_{x(y)}^{n-1}=\mathbb{1} \otimes \ldots \otimes \mathbb{1} \otimes R_{x(y)} $,  connects the relative phases as follows:
\begin{equation}
\begin{split}
    \phi_{i+1}-\phi_{i}=&\,\arctan\left(\frac{-P^x_i+\frac{1}{2}(P_i+P_{i+1})}{P^y_i-\frac{1}{2}(P_i+P_{i+1})}\right) \\
    &i=0,2,4,\ldots,2^n-2\,.
\end{split}
\end{equation}
Employing the two rotations at $j$-th step, one obtains the relatives phases as described in decision tree of Fig.~\ref{fig:pyramid_relative_phases}.\\
One could improve this method by implementing the classical shadow protocol that reduces the number of employed measurements \cite{huang2020predicting,koh2022classical}.
\subsection{Reinitializing operator}
With the presented tomography, we can identify the multiqubit state. After doing so and evaluating the state of the previous time step, we must reinitialize that state to continue simulating the real-time evolution. This section will discuss how we build the reinitializing operator from a theoretical point of view.

We want to build an operator $U_{reinit}$ such that it acts on $\ket{0}$, the initial quantum state, to yield the identified state, $\ket{\psi}$. 
We start by rotating the default state $\ket{0}$ to that with the correct probability distribution. 
One can demonstrate that the following rotation of Eq.~\eqref{eq:tomo_reini_1}, with $\theta_1\,=\,2\,\arcsin(\sqrt{P_1})$, rotates the state $\ket{0}$ to a new one with the correct probability for the state $\ket{1}$, $\ket{\psi_1}=(\cos(\frac{\theta_1}{2}),\, \sin(\frac{\theta_1}{2}),\,0,\,..)^T$.\\
\begin{equation}
  R_1\,=\,  \begin{pmatrix}
    \cos(\frac{\theta_1}{2}) & -\sin(\frac{\theta_1}{2}) & 0 &0 &\ldots&0\\
    \sin(\frac{\theta_1}{2}) & \cos(\frac{\theta_1}{2}) & 0 & 0 &\ldots&0\\
    0 & 0 & 1 & 0 &\ldots& 0\\
    0 & 0 &0 & 1 &\ldots& 0\\
    \ldots & \ldots &\ldots & \ldots & \ldots\\
    0 & 0 &0 & 0 &\ldots& 1\\
    \end{pmatrix}\,. \label{eq:tomo_reini_1}
\end{equation}

To obtain the correct probability distribution for the state $\ket{2}$, we apply the following rotation matrix,
\begin{equation}
    R_2\,=\,  \begin{pmatrix}
    \cos(\frac{\theta_2}{2}) &0  &-\sin(\frac{\theta_2}{2})  &0 &\ldots&0\\
    0 & 1 & 0 & 0 \ldots& 0\\
    \sin(\frac{\theta_2}{2}) & 0& \cos(\frac{\theta_2}{2})  & 0 &\ldots&0\\
    0 & 0 &0 & 1 &\ldots& 0\\
    \ldots & \ldots & \ldots & \ldots &\ldots &\ldots\\
    0 & 0 &0 & 0 &\ldots& 1\\
    \end{pmatrix}\,.
\end{equation}
This moves $\ket{\psi_1}$ to $\ket{\psi_2}=(\cos(\frac{\theta_1}{2})\cos(\frac{\theta_2}{2}),\, \sin(\frac{\theta_1}{2}),\, \cos(\frac{\theta_1}{2})\sin(\frac{\theta_2}{2}),0,\ldots)^T$. To obtain the probability for the second state equal to $P_2$, we choose $\theta_2\,=\,2\,\arcsin\left(\frac{\sqrt{P_2}}{\cos(\frac{\theta_1}{2})}\right)$.

Iterating this algorithm for the state $k$, we $y$-rotate the $\ket{k-1}$ and $\ket{k}$ states by an angle
\begin{equation}
\theta_k\,=\,2\,\arcsin\left(\frac{\sqrt{P_k}}{\prod_i^{k-1} \cos(\frac{\theta_i}{2})}\right)\,.
\end{equation}

The general rotation matrix that moves the state $\ket{0}$ to the state $(\sqrt{P_0},\sqrt{P_1},\ldots,\sqrt{P_N})^T$ is obtained from 
\begin{equation}
    R_{TOT}= R_N\,R_{N-1}\,R_{N-2} \ldots\,R_{2}\,R_1\,R_0\,. \label{eq:tomography_rotationtot}
\end{equation}

To obtain the desired state, we must incorporate the relative phase information. To do so, we apply the following phase gate after $R_{TOT}$:
\begin{equation}
    Ph=\begin{pmatrix}
    1 & 0 & 0 & \ldots & 0\\
    0 & e^{i\,\phi_1} & 0 & \ldots & 0\\
    0 & 0 & e^{i\,\phi_2} & \ldots & 0\\
    \ldots & \ldots& \ldots& \ldots&\ldots\\
    0 & 0 & 0 & \ldots & e^{i\,\phi_N}\\
    \end{pmatrix}\,. \label{eq:tomography_phase}
\end{equation}
This gate implements the proper phases to the state $R_{TOT}\,\ket{0}$.
Ultimately, our final reinitializing operator is given by 
\begin{equation}
   U_{reinit}\,=\,Ph\,R_{TOT}\,, \label{eq:reinitializing}
\end{equation}
where $R_{TOT}$ and $Ph$ are given by Eqs. \eqref{eq:tomography_rotationtot} and \eqref{eq:tomography_phase}, respectively.

Applying this presented algorithm, one can compute a unitary operator whose action reinitializes the state. We use this algorithm for obtaining its matrix form, and we compile it with the \textit{Qiskit} functions for application in the experimental simulation.

\section{JUSTIFICATION OF SPIN INDEPENDENCE FROM THE CLASSICAL SPATIAL COMPONENT \label{app:spin_proof}}

In this section, we will motivate and justify simulating the spatial dynamics using only the spin independent Hamiltonian while neglecting the spin dependent term. This was previously proved in Ref.~\cite{Holland-2020}.

To do so, we work in the interaction picture, where the full nuclear Hamiltonian is written as a sum of a free Hamiltonian, the independent-spin part, and an interaction Hamiltonian, the spin-dependent potential. The spin dependent real time evolution in this picture is given by:
\begin{equation}
    U^I_{\rm SD}(t)= \exp\left[-i\,t\,V^I_{\rm SD}(t)\right]\,,
\end{equation}
where the upper letter $I$($S$) indicates this operator works in Interaction (Schr\"{o}dinger) picture.

The $V^I_{\rm SD}(t)$ operator can be rewritten as follows:
\begin{equation}
    V^I_{\rm SD}(t)\,=\,e^{i t H^S_{\rm SI}}\, V^S_{\rm SD}\,e^{-i t H^S_{\rm SI}}\,,
\end{equation}
where in right-hand side of the equation, we work in Schr\"{o}dinger picture.

Therefore, $ U^I_{\rm SD}(t)$ becomes
\begin{equation}
    U^I_{\rm SD}(t)= \exp\left[-i\,t\,e^{i t H^S_{\rm SI}}\, V^S_{\rm SD}(r) \,e^{-i t H^S_{\rm SI}}\right]\,,
\end{equation}

The spin dependent potential is diagonal for the spatial components, therefore, we can write  $ U^I_{\rm SD}(t)$ as:
\begin{equation}
\begin{split}
    U^I_{\rm SD}(t)&= \exp\left[-i\,t\, V^S_{\rm SD}\left(r(t)\right) \,\right]\\
\end{split}
\end{equation}
where $r(t)\,=\,e^{i t H^S_{\rm SI}}\,\ket{r}\bra{r}\,e^{-i t H^S_{\rm SI}}$. In our case, it is computed classically from the Newton equation.

Therefore, under the assumption that the semiclassical equations of motion for the coordinates are an appropriate approximation, we have proven that the spin dynamics do not affect the spatial dynamics. This justifies the methodology employed in this algorithm.

\section{CIRCUIT QED BACKGROUND\label{app:hardware}}
The spin evolution was carried out experimentally using a superconducting quantum processor at the AQT, operated using the principles of circuit quantum electrodynamics (QED) \cite{cqed_Blais, cqed_Wallraff}. The processor includes floating, fixed frequency transmon qubits, each coupled to their nearest neighbor through a fixed resonator. A transmon, consisting of a capacitively shunted Josephson junction, can be considered an anharmonic oscillator, whereby its lowest two levels are independently addressable \cite{transmon}. It is also possible to controllably address higher levels of a transmon, making it usable as a qutrit \cite{qutrit_Wallraff, qutrit_Goss}.

Qubit state readout was performed in a dispersive manner \cite{single_dispersive_readout_Wallraff, single_dispersive_readout_Mallet, single_dispersive_readout_Walter}. In this regime, the qubit is coupled to a resonator whose frequency is dependent upon the qubit state; therefore, probing the resonator yields information about the qubit state. In this case, the resonators are superconducting, quarter-wavelength ($\lambda/4$) coplanar waveguides (CPW). Individual readout resonators are coupled to a central bus to facilitate multiplexed readout in reflection, where the bus also acts as a wideband Purcell filter encapsulating all readout resonators \cite{Bandpass_purcell_exp, Bandpass_purcell_theory}. Two-qubit gate operations were performed using a CZ gate based on the differential AC Stark shift \cite{diff-AC-Stark}. 

\section{ERROR MITIGATION STRATEGIES}\label{app:Err_Mitigation}
In this section, we discuss in further depth the error mitigation strategies applied to improve the algorithm's experimental performance.

\subsection{Readout Correction}
Common errors that reduce the fidelity of a quantum algorithm include state preparation and measurement (SPAM) errors in addition to gate noise. In the case that the relative error rates $\epsilon_i$ satisfy
\begin{equation}
    \epsilon_\text{Readout} < \epsilon_\text{X gate} + \epsilon_\text{State preparation}\,,
\end{equation}
where an X gate is a rotation by $\pi$ about the $x$-axis of the Bloch sphere, one can perform a simple calibration scheme to help correct readout errors for each qubit. The routine is part of Quantum Benchmark's True-Q\texttrademark \ package \cite{True-Q}, where in broad terms its strategy is to:
\begin{enumerate}
    \item Perform a readout calibration (RCAL) experiment:
    \begin{itemize}
        \item Apply an identity $I$ gate, followed by measurement.
        \item Apply an $X$ gate, followed by measurement.
    \end{itemize}
    \item Calculate the deviation of expected versus observed measurement results, characterized in a measurement confusion matrix.
\end{enumerate}
To correct the raw bit string results of a circuit, the measured confusion matrix for each qubit is inverted and contracted onto the corresponding qubit index in the measurement results. The calibration step (1) is constant in the number of qubits, since the gates on all qubits can be performed in parallel, therefore requiring only two circuits for $n$ qubits. However, the inversion step has exponential scaling in the number of qubits, and is therefore not scalable.

\subsection{Randomized Compiling (RC)}
Randomized compiling is a scalable error mitigation method applied to a circuit that seeks to translate coherent error into stochastic error, while not changing the logical circuit or increasing its depth ~\cite{RC_original, RC_Akel}. This is done by inserting random virtual twirling gates and their corresponding inverting gates into the circuit, here sampled from the Pauli group, such that the final circuit constitutes the same unitary operation but with different single-qubit gates. This procedure is repeated over many instances, with the end result being a collection of random circuits of the same depth that are logically equivalent.\\

Repeatedly running each circuit in this collection will yield a statistical distribution of results where coherent error from the original circuit has been tailored into stochastic error. While coherent errors can build up quadratically with circuit depth, stochastic errors only build up linearly with circuit depth. Because most circuits in this study were measured with 10,000 repetitions, the same number of measurements was kept constant for experiments using RC. Therefore, using $20$ randomizations under RC, each of these $20$ circuits was measured $500$ times.\\

\begin{figure}
    \centering
    \includegraphics[scale=0.38]{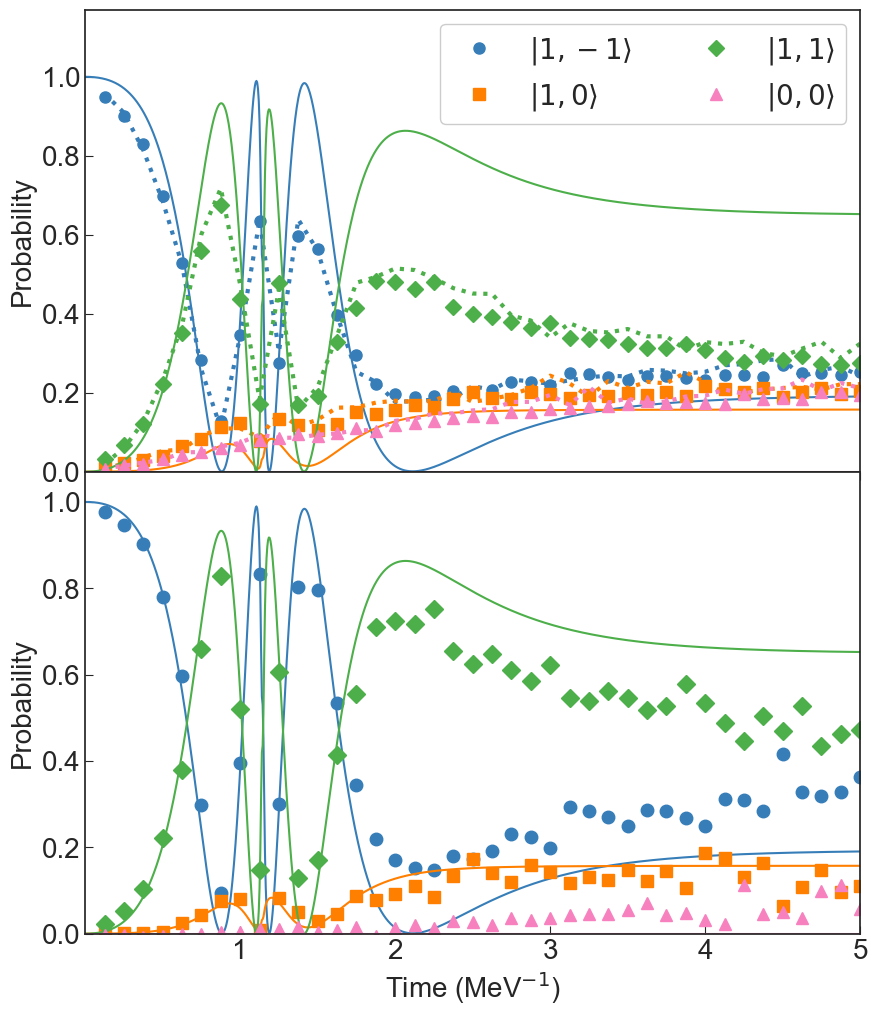}
    \caption{Applications of RC and state purification. The top panel (a) displays the data after application of only RC, with the dashed lines indicating the simulated results under a fully depolarizing noise model using the measured process fidelity. The bottom panel (b) displays the data after state purification, where the effects of decoherence have been further mitigated.}
    \label{fig:RC_Purificiation_RO_corr_40circ}
\end{figure}

\subsection{State purification}

While the initial state of a system may be pure, over time due to decoherence and other errors, the state will become mixed. In a geometric sense, this process can be represented as the initial generalized Bloch vector having unit length, and as the algorithm progresses in time, it becomes shorter and shrinks towards the origin. In this experiment, state purification entails taking a partially mixed quantum state and transforming it to a pure one. This can be thought of as renormalizing the length of the Bloch vector back to unit length, or on the Bloch sphere's surface, which can be accomplished by estimating the extent that the Bloch vector length has reduced.\\

While more exact forms of state purification may rely upon state tomography protocols to extract the length of the Bloch vector from a density matrix, in this experimental protocol, we use a simplified version that is much less resource-intensive and more scalable. Under RC, we assume that an approximately depolarizing noise model can fit the observed results, which can be verified by introducing depolarizing noise into the simulated results, as shown in Fig. \ref{fig:RC_Purificiation_RO_corr_40circ}(a). The depolarizing noise channel can be represented as a map \cite{Nielsen_Chuang}
\begin{equation}
    \epsilon(\rho) = \lambda \rho + (1-\lambda) \frac{I}{2^n}\,,
\end{equation}
where $(1-\lambda)$ can be considered the probability that the density matrix $\rho$ is mapped onto the maximally mixed state, so $0 \leq \lambda \leq 1$, and $n$ is the number of qubits. The factor $\lambda$ is the effective depolarizing parameter, which is linearly related to the process infidelity of the two-qubit CZ gate, which can be measured via cycle benchmarking \cite{cycle_benchmarking}. Using it, we can renormalize the obtained Pauli expectation values by a factor $\lambda^N$, where $N$ is the number of two-qubit gates that appear in the circuit. This renormalization is performed as a postprocessing step using only the benchmarked gate process fidelity, so once an estimate for $\lambda$ is obtained, its application for this state purification is computationally inexpensive.

\subsection{Circuit compression}
As an additional error mitigation method, we applied the Cartan, or KAK, decomposition to decompose any two-qubit unitary into a circuit with at most three $CNOT$ gates \cite{Cartan}. This method is not scalable to larger numbers of qubits, but it is viable in this experiment using only two qubits. Thus, it was used to evaluate the limits on the potential accuracy of the experimental algorithm, although we acknowledge that such circuit compression becomes expensive or unfeasible as the number of qubits increases. Circuit compression also enables validation of the algorithm's procedure at finer time steps for the same amount of time. 

Fig. \ref{fig:RC_Purificiation_RO_corr_40circ} displays the results of running the algorithm when splitting it into 40 time steps. In the top panel, RC is applied, and one can observe all state probabilities converging to each other towards the end. Purifying the results will improve the algorithm's accuracy, although it too cannot sufficiently mitigate against the errors at higher circuit depths. Thus, if wanting to run the algorithm at a finer resolution in time, circuit compression had to be used. This is because it results in all time steps only requiring at most three two-qubit gates, such that the circuit depth no longer scales linearly with time.


\begin{thebibliography}{64}%
\makeatletter
\providecommand \@ifxundefined [1]{%
 \@ifx{#1\undefined}
}%
\providecommand \@ifnum [1]{%
 \ifnum #1\expandafter \@firstoftwo
 \else \expandafter \@secondoftwo
 \fi
}%
\providecommand \@ifx [1]{%
 \ifx #1\expandafter \@firstoftwo
 \else \expandafter \@secondoftwo
 \fi
}%
\providecommand \natexlab [1]{#1}%
\providecommand \enquote  [1]{``#1''}%
\providecommand \bibnamefont  [1]{#1}%
\providecommand \bibfnamefont [1]{#1}%
\providecommand \citenamefont [1]{#1}%
\providecommand \href@noop [0]{\@secondoftwo}%
\providecommand \href [0]{\begingroup \@sanitize@url \@href}%
\providecommand \@href[1]{\@@startlink{#1}\@@href}%
\providecommand \@@href[1]{\endgroup#1\@@endlink}%
\providecommand \@sanitize@url [0]{\catcode `\\12\catcode `\$12\catcode
  `\&12\catcode `\#12\catcode `\^12\catcode `\_12\catcode `\%12\relax}%
\providecommand \@@startlink[1]{}%
\providecommand \@@endlink[0]{}%
\providecommand \url  [0]{\begingroup\@sanitize@url \@url }%
\providecommand \@url [1]{\endgroup\@href {#1}{\urlprefix }}%
\providecommand \urlprefix  [0]{URL }%
\providecommand \Eprint [0]{\href }%
\providecommand \doibase [0]{http://dx.doi.org/}%
\providecommand \selectlanguage [0]{\@gobble}%
\providecommand \bibinfo  [0]{\@secondoftwo}%
\providecommand \bibfield  [0]{\@secondoftwo}%
\providecommand \translation [1]{[#1]}%
\providecommand \BibitemOpen [0]{}%
\providecommand \bibitemStop [0]{}%
\providecommand \bibitemNoStop [0]{.\EOS\space}%
\providecommand \EOS [0]{\spacefactor3000\relax}%
\providecommand \BibitemShut  [1]{\csname bibitem#1\endcsname}%
\let\auto@bib@innerbib\@empty
\bibitem [{\citenamefont {Nunes}\ \emph {et~al.}(2020)\citenamefont {Nunes},
  \citenamefont {Potel}, \citenamefont {Poxon-Pearson},\ and\ \citenamefont
  {Cizewski}}]{doi:10.1146/annurev-nucl-020620-063734}%
  \BibitemOpen
  \bibfield  {author} {\bibinfo {author} {\bibfnamefont {F.}~\bibnamefont
  {Nunes}}, \bibinfo {author} {\bibfnamefont {G.}~\bibnamefont {Potel}},
  \bibinfo {author} {\bibfnamefont {T.}~\bibnamefont {Poxon-Pearson}}, \ and\
  \bibinfo {author} {\bibfnamefont {J.}~\bibnamefont {Cizewski}},\ }\href
  {\doibase 10.1146/annurev-nucl-020620-063734} {\bibfield  {journal} {\bibinfo
   {journal} {Annual Review of Nuclear and Particle Science}\ }\textbf
  {\bibinfo {volume} {70}},\ \bibinfo {pages} {147} (\bibinfo {year}
  {2020})}\BibitemShut {NoStop}%
\bibitem [{\citenamefont {Wiescher}\ \emph {et~al.}(2012)\citenamefont
  {Wiescher}, \citenamefont {K\"{a}ppeler},\ and\ \citenamefont
  {Langanke}}]{doi:10.1146/annurev-astro-081811-125543}%
  \BibitemOpen
  \bibfield  {author} {\bibinfo {author} {\bibfnamefont {M.}~\bibnamefont
  {Wiescher}}, \bibinfo {author} {\bibfnamefont {F.}~\bibnamefont
  {K\"{a}ppeler}}, \ and\ \bibinfo {author} {\bibfnamefont {K.}~\bibnamefont
  {Langanke}},\ }\href {\doibase 10.1146/annurev-astro-081811-125543}
  {\bibfield  {journal} {\bibinfo  {journal} {Annual Review of Astronomy and
  Astrophysics}\ }\textbf {\bibinfo {volume} {50}},\ \bibinfo {pages} {165}
  (\bibinfo {year} {2012})}\BibitemShut {NoStop}%
\bibitem [{\citenamefont {Adelberger}\ \emph {et~al.}(2011)\citenamefont
  {Adelberger}, \citenamefont {Garc\'{\i}a}, \citenamefont {Robertson},
  \citenamefont {Snover}, \citenamefont {Balantekin}, \citenamefont {Heeger},
  \citenamefont {Ramsey-Musolf}, \citenamefont {Bemmerer}, \citenamefont
  {Junghans}, \citenamefont {Bertulani}, \citenamefont {Chen}, \citenamefont
  {Costantini}, \citenamefont {Prati}, \citenamefont {Couder}, \citenamefont
  {Uberseder}, \citenamefont {Wiescher}, \citenamefont {Cyburt}, \citenamefont
  {Davids}, \citenamefont {Freedman}, \citenamefont {Gai}, \citenamefont
  {Gazit}, \citenamefont {Gialanella}, \citenamefont {Imbriani}, \citenamefont
  {Greife}, \citenamefont {Hass}, \citenamefont {Haxton}, \citenamefont
  {Itahashi}, \citenamefont {Kubodera}, \citenamefont {Langanke}, \citenamefont
  {Leitner}, \citenamefont {Leitner}, \citenamefont {Vetter}, \citenamefont
  {Winslow}, \citenamefont {Marcucci}, \citenamefont {Motobayashi},
  \citenamefont {Mukhamedzhanov}, \citenamefont {Tribble}, \citenamefont
  {Nollett}, \citenamefont {Nunes}, \citenamefont {Park}, \citenamefont
  {Parker}, \citenamefont {Schiavilla}, \citenamefont {Simpson}, \citenamefont
  {Spitaleri}, \citenamefont {Strieder}, \citenamefont {Trautvetter},
  \citenamefont {Suemmerer},\ and\ \citenamefont {Typel}}]{RevModPhys_83_195}%
  \BibitemOpen
  \bibfield  {author} {\bibinfo {author} {\bibfnamefont {E.~G.}\ \bibnamefont
  {Adelberger}}, \bibinfo {author} {\bibfnamefont {A.}~\bibnamefont
  {Garc\'{\i}a}}, \bibinfo {author} {\bibfnamefont {R.}~\bibnamefont
  {Robertson}}, \bibinfo {author} {\bibfnamefont {K.~A.}\ \bibnamefont
  {Snover}}, \bibinfo {author} {\bibfnamefont {A.~B.}\ \bibnamefont
  {Balantekin}}, \bibinfo {author} {\bibfnamefont {K.}~\bibnamefont {Heeger}},
  \bibinfo {author} {\bibfnamefont {M.~J.}\ \bibnamefont {Ramsey-Musolf}},
  \bibinfo {author} {\bibfnamefont {D.}~\bibnamefont {Bemmerer}}, \bibinfo
  {author} {\bibfnamefont {A.}~\bibnamefont {Junghans}}, \bibinfo {author}
  {\bibfnamefont {C.~A.}\ \bibnamefont {Bertulani}}, \bibinfo {author}
  {\bibfnamefont {J.~W.}\ \bibnamefont {Chen}}, \bibinfo {author}
  {\bibfnamefont {H.}~\bibnamefont {Costantini}}, \bibinfo {author}
  {\bibfnamefont {P.}~\bibnamefont {Prati}}, \bibinfo {author} {\bibfnamefont
  {M.}~\bibnamefont {Couder}}, \bibinfo {author} {\bibfnamefont
  {E.}~\bibnamefont {Uberseder}}, \bibinfo {author} {\bibfnamefont
  {M.}~\bibnamefont {Wiescher}}, \bibinfo {author} {\bibfnamefont
  {R.}~\bibnamefont {Cyburt}}, \bibinfo {author} {\bibfnamefont
  {B.}~\bibnamefont {Davids}}, \bibinfo {author} {\bibfnamefont {S.~J.}\
  \bibnamefont {Freedman}}, \bibinfo {author} {\bibfnamefont {M.}~\bibnamefont
  {Gai}}, \bibinfo {author} {\bibfnamefont {D.}~\bibnamefont {Gazit}}, \bibinfo
  {author} {\bibfnamefont {L.}~\bibnamefont {Gialanella}}, \bibinfo {author}
  {\bibfnamefont {G.}~\bibnamefont {Imbriani}}, \bibinfo {author}
  {\bibfnamefont {U.}~\bibnamefont {Greife}}, \bibinfo {author} {\bibfnamefont
  {M.}~\bibnamefont {Hass}}, \bibinfo {author} {\bibfnamefont {W.~C.}\
  \bibnamefont {Haxton}}, \bibinfo {author} {\bibfnamefont {T.}~\bibnamefont
  {Itahashi}}, \bibinfo {author} {\bibfnamefont {K.}~\bibnamefont {Kubodera}},
  \bibinfo {author} {\bibfnamefont {K.}~\bibnamefont {Langanke}}, \bibinfo
  {author} {\bibfnamefont {D.}~\bibnamefont {Leitner}}, \bibinfo {author}
  {\bibfnamefont {M.}~\bibnamefont {Leitner}}, \bibinfo {author} {\bibfnamefont
  {P.}~\bibnamefont {Vetter}}, \bibinfo {author} {\bibfnamefont
  {L.}~\bibnamefont {Winslow}}, \bibinfo {author} {\bibfnamefont {L.~E.}\
  \bibnamefont {Marcucci}}, \bibinfo {author} {\bibfnamefont {T.}~\bibnamefont
  {Motobayashi}}, \bibinfo {author} {\bibfnamefont {A.}~\bibnamefont
  {Mukhamedzhanov}}, \bibinfo {author} {\bibfnamefont {R.~E.}\ \bibnamefont
  {Tribble}}, \bibinfo {author} {\bibfnamefont {K.~M.}\ \bibnamefont
  {Nollett}}, \bibinfo {author} {\bibfnamefont {F.~M.}\ \bibnamefont {Nunes}},
  \bibinfo {author} {\bibfnamefont {T.-S.}\ \bibnamefont {Park}}, \bibinfo
  {author} {\bibfnamefont {P.~D.}\ \bibnamefont {Parker}}, \bibinfo {author}
  {\bibfnamefont {R.}~\bibnamefont {Schiavilla}}, \bibinfo {author}
  {\bibfnamefont {E.~C.}\ \bibnamefont {Simpson}}, \bibinfo {author}
  {\bibfnamefont {C.}~\bibnamefont {Spitaleri}}, \bibinfo {author}
  {\bibfnamefont {F.}~\bibnamefont {Strieder}}, \bibinfo {author}
  {\bibfnamefont {H.-P.}\ \bibnamefont {Trautvetter}}, \bibinfo {author}
  {\bibfnamefont {K.}~\bibnamefont {Suemmerer}}, \ and\ \bibinfo {author}
  {\bibfnamefont {S.}~\bibnamefont {Typel}},\ }\href {\doibase
  10.1103/RevModPhys.83.195} {\bibfield  {journal} {\bibinfo  {journal} {Rev.
  Mod. Phys.}\ }\textbf {\bibinfo {volume} {83}},\ \bibinfo {pages} {195}
  (\bibinfo {year} {2011})}\BibitemShut {NoStop}%
\bibitem [{\citenamefont {Giustino}(2017)}]{RevModPhys_89_015003}%
  \BibitemOpen
  \bibfield  {author} {\bibinfo {author} {\bibfnamefont {F.}~\bibnamefont
  {Giustino}},\ }\href {\doibase 10.1103/RevModPhys.89.015003} {\bibfield
  {journal} {\bibinfo  {journal} {Rev. Mod. Phys.}\ }\textbf {\bibinfo {volume}
  {89}},\ \bibinfo {pages} {015003} (\bibinfo {year} {2017})}\BibitemShut
  {NoStop}%
\bibitem [{\citenamefont {Quan}\ \emph {et~al.}(2021)\citenamefont {Quan},
  \citenamefont {Yue},\ and\ \citenamefont
  {Liao}}]{doi:10.1080/15567265.2021.1902441}%
  \BibitemOpen
  \bibfield  {author} {\bibinfo {author} {\bibfnamefont {Y.}~\bibnamefont
  {Quan}}, \bibinfo {author} {\bibfnamefont {S.}~\bibnamefont {Yue}}, \ and\
  \bibinfo {author} {\bibfnamefont {B.}~\bibnamefont {Liao}},\ }\href {\doibase
  10.1080/15567265.2021.1902441} {\bibfield  {journal} {\bibinfo  {journal}
  {Nanoscale and Microscale Thermophysical Engineering}\ }\textbf {\bibinfo
  {volume} {25}},\ \bibinfo {pages} {73} (\bibinfo {year} {2021})}\BibitemShut
  {NoStop}%
\bibitem [{\citenamefont {Huang}\ \emph
  {et~al.}(2020{\natexlab{a}})\citenamefont {Huang}, \citenamefont {Wu},
  \citenamefont {Fan},\ and\ \citenamefont {Zhu}}]{huang2020superconducting}%
  \BibitemOpen
  \bibfield  {author} {\bibinfo {author} {\bibfnamefont {H.-L.}\ \bibnamefont
  {Huang}}, \bibinfo {author} {\bibfnamefont {D.}~\bibnamefont {Wu}}, \bibinfo
  {author} {\bibfnamefont {D.}~\bibnamefont {Fan}}, \ and\ \bibinfo {author}
  {\bibfnamefont {X.}~\bibnamefont {Zhu}},\ }\href {\doibase
  10.1007/s11432-020-2881-9} {\bibfield  {journal} {\bibinfo  {journal} {Sci.
  China Inf. Sci.}\ }\textbf {\bibinfo {volume} {63}},\ \bibinfo {pages}
  {180501} (\bibinfo {year} {2020}{\natexlab{a}})}\BibitemShut {NoStop}%
\bibitem [{\citenamefont {Bruzewicz}\ \emph {et~al.}(2019)\citenamefont
  {Bruzewicz}, \citenamefont {Chiaverini}, \citenamefont {McConnell},\ and\
  \citenamefont {Sage}}]{bruzewicz2019trapped}%
  \BibitemOpen
  \bibfield  {author} {\bibinfo {author} {\bibfnamefont {C.~D.}\ \bibnamefont
  {Bruzewicz}}, \bibinfo {author} {\bibfnamefont {J.}~\bibnamefont
  {Chiaverini}}, \bibinfo {author} {\bibfnamefont {R.}~\bibnamefont
  {McConnell}}, \ and\ \bibinfo {author} {\bibfnamefont {J.~M.}\ \bibnamefont
  {Sage}},\ }\href {\doibase 10.1063/1.5088164} {\bibfield  {journal} {\bibinfo
   {journal} {Applied Physics Reviews}\ }\textbf {\bibinfo {volume} {6}},\
  \bibinfo {pages} {021314} (\bibinfo {year} {2019})}\BibitemShut {NoStop}%
\bibitem [{\citenamefont {Henriet}\ \emph {et~al.}(2020)\citenamefont
  {Henriet}, \citenamefont {Beguin}, \citenamefont {Signoles}, \citenamefont
  {Lahaye}, \citenamefont {Browaeys}, \citenamefont {Reymond},\ and\
  \citenamefont {Jurczak}}]{henriet2020quantum}%
  \BibitemOpen
  \bibfield  {author} {\bibinfo {author} {\bibfnamefont {L.}~\bibnamefont
  {Henriet}}, \bibinfo {author} {\bibfnamefont {L.}~\bibnamefont {Beguin}},
  \bibinfo {author} {\bibfnamefont {A.}~\bibnamefont {Signoles}}, \bibinfo
  {author} {\bibfnamefont {T.}~\bibnamefont {Lahaye}}, \bibinfo {author}
  {\bibfnamefont {A.}~\bibnamefont {Browaeys}}, \bibinfo {author}
  {\bibfnamefont {G.-O.}\ \bibnamefont {Reymond}}, \ and\ \bibinfo {author}
  {\bibfnamefont {C.}~\bibnamefont {Jurczak}},\ }\href {\doibase
  10.22331/q-2020-09-21-327} {\bibfield  {journal} {\bibinfo  {journal}
  {Quantum}\ }\textbf {\bibinfo {volume} {4}},\ \bibinfo {pages} {327}
  (\bibinfo {year} {2020})}\BibitemShut {NoStop}%
\bibitem [{\citenamefont {Jordan}\ \emph {et~al.}(2012)\citenamefont {Jordan},
  \citenamefont {Lee},\ and\ \citenamefont
  {Preskill}}]{Jordan-Lee-Preskill-2012}%
  \BibitemOpen
  \bibfield  {author} {\bibinfo {author} {\bibfnamefont {S.~P.}\ \bibnamefont
  {Jordan}}, \bibinfo {author} {\bibfnamefont {K.~S.~M.}\ \bibnamefont {Lee}},
  \ and\ \bibinfo {author} {\bibfnamefont {J.}~\bibnamefont {Preskill}},\
  }\href {\doibase 10.1126/science.1217069} {\bibfield  {journal} {\bibinfo
  {journal} {Science}\ }\textbf {\bibinfo {volume} {336}},\ \bibinfo {pages}
  {1130} (\bibinfo {year} {2012})}\BibitemShut {NoStop}%
\bibitem [{\citenamefont {Liu}\ \emph {et~al.}(2021)\citenamefont {Liu},
  \citenamefont {Sun},\ and\ \citenamefont {Yuan}}]{liu2021towards}%
  \BibitemOpen
  \bibfield  {author} {\bibinfo {author} {\bibfnamefont {J.}~\bibnamefont
  {Liu}}, \bibinfo {author} {\bibfnamefont {J.}~\bibnamefont {Sun}}, \ and\
  \bibinfo {author} {\bibfnamefont {X.}~\bibnamefont {Yuan}},\ }\href {\doibase
  10.1088/2632-2153/aca06b} {\bibfield  {journal} {\bibinfo  {journal} {Mach.
  Learn.: Sci. Technol.}\ }\textbf {\bibinfo {volume} {3}},\ \bibinfo {pages}
  {045030} (\bibinfo {year} {2021})}\BibitemShut {NoStop}%
\bibitem [{\citenamefont {Peruzzo}\ \emph {et~al.}(2014)\citenamefont
  {Peruzzo}, \citenamefont {McClean}, \citenamefont {Shadbolt}, \citenamefont
  {Yung}, \citenamefont {Zhou}, \citenamefont {Love}, \citenamefont
  {Aspuru-Guzik},\ and\ \citenamefont {O'Brien}}]{Peruzzo-2014}%
  \BibitemOpen
  \bibfield  {author} {\bibinfo {author} {\bibfnamefont {A.}~\bibnamefont
  {Peruzzo}}, \bibinfo {author} {\bibfnamefont {J.}~\bibnamefont {McClean}},
  \bibinfo {author} {\bibfnamefont {P.}~\bibnamefont {Shadbolt}}, \bibinfo
  {author} {\bibfnamefont {M.-H.}\ \bibnamefont {Yung}}, \bibinfo {author}
  {\bibfnamefont {X.-Q.}\ \bibnamefont {Zhou}}, \bibinfo {author}
  {\bibfnamefont {P.~J.}\ \bibnamefont {Love}}, \bibinfo {author}
  {\bibfnamefont {A.}~\bibnamefont {Aspuru-Guzik}}, \ and\ \bibinfo {author}
  {\bibfnamefont {J.~L.}\ \bibnamefont {O'Brien}},\ }\href {\doibase
  10.1038/ncomms5213} {\bibfield  {journal} {\bibinfo  {journal} {Nat.
  Commun.}\ }\textbf {\bibinfo {volume} {5}},\ \bibinfo {pages} {4213}
  (\bibinfo {year} {2014})}\BibitemShut {NoStop}%
\bibitem [{\citenamefont {McClean}\ \emph {et~al.}(2016)\citenamefont
  {McClean}, \citenamefont {Romero}, \citenamefont {Babbush},\ and\
  \citenamefont {Aspuru-Guzik}}]{mcclean2016theory}%
  \BibitemOpen
  \bibfield  {author} {\bibinfo {author} {\bibfnamefont {J.~R.}\ \bibnamefont
  {McClean}}, \bibinfo {author} {\bibfnamefont {J.}~\bibnamefont {Romero}},
  \bibinfo {author} {\bibfnamefont {R.}~\bibnamefont {Babbush}}, \ and\
  \bibinfo {author} {\bibfnamefont {A.}~\bibnamefont {Aspuru-Guzik}},\ }\href
  {\doibase 10.1088/1367-2630/18/2/023023} {\bibfield  {journal} {\bibinfo
  {journal} {New J. Phys.}\ }\textbf {\bibinfo {volume} {18}},\ \bibinfo
  {pages} {023023} (\bibinfo {year} {2016})}\BibitemShut {NoStop}%
\bibitem [{\citenamefont {Arute}\ \emph {et~al.}(2020)\citenamefont {Arute},
  \citenamefont {Arya}, \citenamefont {Babbush}, \citenamefont {Bacon},
  \citenamefont {Bardin}, \citenamefont {Barends}, \citenamefont {Boixo},
  \citenamefont {Broughton}, \citenamefont {Buckley} \emph
  {et~al.}}]{google2020hartree}%
  \BibitemOpen
  \bibfield  {author} {\bibinfo {author} {\bibfnamefont {F.}~\bibnamefont
  {Arute}}, \bibinfo {author} {\bibfnamefont {K.}~\bibnamefont {Arya}},
  \bibinfo {author} {\bibfnamefont {R.}~\bibnamefont {Babbush}}, \bibinfo
  {author} {\bibfnamefont {D.}~\bibnamefont {Bacon}}, \bibinfo {author}
  {\bibfnamefont {J.~C.}\ \bibnamefont {Bardin}}, \bibinfo {author}
  {\bibfnamefont {R.}~\bibnamefont {Barends}}, \bibinfo {author} {\bibfnamefont
  {S.}~\bibnamefont {Boixo}}, \bibinfo {author} {\bibfnamefont
  {M.}~\bibnamefont {Broughton}}, \bibinfo {author} {\bibfnamefont {B.~B.}\
  \bibnamefont {Buckley}},  \emph {et~al.},\ }\href {\doibase
  10.1126/science.abb9811} {\bibfield  {journal} {\bibinfo  {journal}
  {Science}\ }\textbf {\bibinfo {volume} {369}},\ \bibinfo {pages} {1084}
  (\bibinfo {year} {2020})}\BibitemShut {NoStop}%
\bibitem [{\citenamefont {Motta}\ \emph {et~al.}(2020)\citenamefont {Motta},
  \citenamefont {Sun}, \citenamefont {Tan}, \citenamefont {O’Rourke},
  \citenamefont {Ye}, \citenamefont {Minnich}, \citenamefont {Brand{\~a}o},\
  and\ \citenamefont {Chan}}]{motta2020determining}%
  \BibitemOpen
  \bibfield  {author} {\bibinfo {author} {\bibfnamefont {M.}~\bibnamefont
  {Motta}}, \bibinfo {author} {\bibfnamefont {C.}~\bibnamefont {Sun}}, \bibinfo
  {author} {\bibfnamefont {A.~T.}\ \bibnamefont {Tan}}, \bibinfo {author}
  {\bibfnamefont {M.~J.}\ \bibnamefont {O’Rourke}}, \bibinfo {author}
  {\bibfnamefont {E.}~\bibnamefont {Ye}}, \bibinfo {author} {\bibfnamefont
  {A.~J.}\ \bibnamefont {Minnich}}, \bibinfo {author} {\bibfnamefont {F.~G.}\
  \bibnamefont {Brand{\~a}o}}, \ and\ \bibinfo {author} {\bibfnamefont
  {G.~K.-L.}\ \bibnamefont {Chan}},\ }\href {\doibase
  10.1038/s41567-019-0704-4} {\bibfield  {journal} {\bibinfo  {journal} {Nat.
  Phys.}\ }\textbf {\bibinfo {volume} {16}},\ \bibinfo {pages} {205} (\bibinfo
  {year} {2020})}\BibitemShut {NoStop}%
\bibitem [{\citenamefont {McArdle}\ \emph {et~al.}(2019)\citenamefont
  {McArdle}, \citenamefont {Jones}, \citenamefont {Endo}, \citenamefont {Li},
  \citenamefont {Benjamin},\ and\ \citenamefont
  {Yuan}}]{mcardle2019variational}%
  \BibitemOpen
  \bibfield  {author} {\bibinfo {author} {\bibfnamefont {S.}~\bibnamefont
  {McArdle}}, \bibinfo {author} {\bibfnamefont {T.}~\bibnamefont {Jones}},
  \bibinfo {author} {\bibfnamefont {S.}~\bibnamefont {Endo}}, \bibinfo {author}
  {\bibfnamefont {Y.}~\bibnamefont {Li}}, \bibinfo {author} {\bibfnamefont
  {S.~C.}\ \bibnamefont {Benjamin}}, \ and\ \bibinfo {author} {\bibfnamefont
  {X.}~\bibnamefont {Yuan}},\ }\href {\doibase 10.1038/s41524-019-0187-2}
  {\bibfield  {journal} {\bibinfo  {journal} {npj Quantum Inf.}\ }\textbf
  {\bibinfo {volume} {5}},\ \bibinfo {pages} {75} (\bibinfo {year}
  {2019})}\BibitemShut {NoStop}%
\bibitem [{\citenamefont {Epelbaum}\ \emph {et~al.}(2009)\citenamefont
  {Epelbaum}, \citenamefont {Hammer},\ and\ \citenamefont
  {Mei\ss{}ner}}]{RevModPhys.81.1773}%
  \BibitemOpen
  \bibfield  {author} {\bibinfo {author} {\bibfnamefont {E.}~\bibnamefont
  {Epelbaum}}, \bibinfo {author} {\bibfnamefont {H.-W.}\ \bibnamefont
  {Hammer}}, \ and\ \bibinfo {author} {\bibfnamefont {U.-G.}\ \bibnamefont
  {Mei\ss{}ner}},\ }\href {\doibase 10.1103/RevModPhys.81.1773} {\bibfield
  {journal} {\bibinfo  {journal} {Rev. Mod. Phys.}\ }\textbf {\bibinfo {volume}
  {81}},\ \bibinfo {pages} {1773} (\bibinfo {year} {2009})}\BibitemShut
  {NoStop}%
\bibitem [{\citenamefont {Carlson}\ \emph {et~al.}(2015)\citenamefont
  {Carlson}, \citenamefont {Gandolfi}, \citenamefont {Pederiva}, \citenamefont
  {Pieper}, \citenamefont {Schiavilla}, \citenamefont {Schmidt},\ and\
  \citenamefont {Wiringa}}]{RevModPhys.87.1067}%
  \BibitemOpen
  \bibfield  {author} {\bibinfo {author} {\bibfnamefont {J.}~\bibnamefont
  {Carlson}}, \bibinfo {author} {\bibfnamefont {S.}~\bibnamefont {Gandolfi}},
  \bibinfo {author} {\bibfnamefont {F.}~\bibnamefont {Pederiva}}, \bibinfo
  {author} {\bibfnamefont {S.~C.}\ \bibnamefont {Pieper}}, \bibinfo {author}
  {\bibfnamefont {R.}~\bibnamefont {Schiavilla}}, \bibinfo {author}
  {\bibfnamefont {K.~E.}\ \bibnamefont {Schmidt}}, \ and\ \bibinfo {author}
  {\bibfnamefont {R.~B.}\ \bibnamefont {Wiringa}},\ }\href {\doibase
  10.1103/RevModPhys.87.1067} {\bibfield  {journal} {\bibinfo  {journal} {Rev.
  Mod. Phys.}\ }\textbf {\bibinfo {volume} {87}},\ \bibinfo {pages} {1067}
  (\bibinfo {year} {2015})}\BibitemShut {NoStop}%
\bibitem [{\citenamefont {Trotter}(1959)}]{Trotter-1959}%
  \BibitemOpen
  \bibfield  {author} {\bibinfo {author} {\bibfnamefont {H.~F.}\ \bibnamefont
  {Trotter}},\ }\href {http://www.jstor.org/stable/2033649} {\bibfield
  {journal} {\bibinfo  {journal} {Proc. Amer. Math. Soc.}\ }\textbf {\bibinfo
  {volume} {10}},\ \bibinfo {pages} {545} (\bibinfo {year} {1959})}\BibitemShut
  {NoStop}%
\bibitem [{\citenamefont {Suzuki}(1993)}]{suzuki-1993}%
  \BibitemOpen
  \bibfield  {author} {\bibinfo {author} {\bibfnamefont {M.}~\bibnamefont
  {Suzuki}},\ }\href {\doibase 10.2183/pjab.69.161} {\bibfield  {journal}
  {\bibinfo  {journal} {Proc. Jpn. Acad., Ser. B, Phys.}\ }\textbf {\bibinfo
  {volume} {69}},\ \bibinfo {pages} {161} (\bibinfo {year} {1993})}\BibitemShut
  {NoStop}%
\bibitem [{\citenamefont {Turro}\ \emph {et~al.}(2022)\citenamefont {Turro},
  \citenamefont {Roggero}, \citenamefont {Amitrano}, \citenamefont {Luchi},
  \citenamefont {Wendt}, \citenamefont {Dubois}, \citenamefont {Quaglioni},\
  and\ \citenamefont {Pederiva}}]{Turro2022}%
  \BibitemOpen
  \bibfield  {author} {\bibinfo {author} {\bibfnamefont {F.}~\bibnamefont
  {Turro}}, \bibinfo {author} {\bibfnamefont {A.}~\bibnamefont {Roggero}},
  \bibinfo {author} {\bibfnamefont {V.}~\bibnamefont {Amitrano}}, \bibinfo
  {author} {\bibfnamefont {P.}~\bibnamefont {Luchi}}, \bibinfo {author}
  {\bibfnamefont {K.~A.}\ \bibnamefont {Wendt}}, \bibinfo {author}
  {\bibfnamefont {J.~L.}\ \bibnamefont {Dubois}}, \bibinfo {author}
  {\bibfnamefont {S.}~\bibnamefont {Quaglioni}}, \ and\ \bibinfo {author}
  {\bibfnamefont {F.}~\bibnamefont {Pederiva}},\ }\href {\doibase
  10.1103/PhysRevA.105.022440} {\bibfield  {journal} {\bibinfo  {journal}
  {Phys. Rev. A}\ }\textbf {\bibinfo {volume} {105}},\ \bibinfo {pages}
  {022440} (\bibinfo {year} {2022})}\BibitemShut {NoStop}%
\bibitem [{\citenamefont {Ville}\ \emph {et~al.}(2022)\citenamefont {Ville},
  \citenamefont {Morvan}, \citenamefont {Hashim}, \citenamefont {Naik},
  \citenamefont {Lu}, \citenamefont {Mitchell}, \citenamefont {Kreikebaum},
  \citenamefont {O'Brien}, \citenamefont {Wallman}, \citenamefont {Hincks},
  \citenamefont {Emerson}, \citenamefont {Smith}, \citenamefont {Younis},
  \citenamefont {Iancu}, \citenamefont {Santiago},\ and\ \citenamefont
  {Siddiqi}}]{QITE_JL}%
  \BibitemOpen
  \bibfield  {author} {\bibinfo {author} {\bibfnamefont {J.-L.}\ \bibnamefont
  {Ville}}, \bibinfo {author} {\bibfnamefont {A.}~\bibnamefont {Morvan}},
  \bibinfo {author} {\bibfnamefont {A.}~\bibnamefont {Hashim}}, \bibinfo
  {author} {\bibfnamefont {R.~K.}\ \bibnamefont {Naik}}, \bibinfo {author}
  {\bibfnamefont {M.}~\bibnamefont {Lu}}, \bibinfo {author} {\bibfnamefont
  {B.~K.}\ \bibnamefont {Mitchell}}, \bibinfo {author} {\bibfnamefont {J.~M.}\
  \bibnamefont {Kreikebaum}}, \bibinfo {author} {\bibfnamefont {K.~P.}\
  \bibnamefont {O'Brien}}, \bibinfo {author} {\bibfnamefont {J.~J.}\
  \bibnamefont {Wallman}}, \bibinfo {author} {\bibfnamefont {I.}~\bibnamefont
  {Hincks}}, \bibinfo {author} {\bibfnamefont {J.}~\bibnamefont {Emerson}},
  \bibinfo {author} {\bibfnamefont {E.}~\bibnamefont {Smith}}, \bibinfo
  {author} {\bibfnamefont {E.}~\bibnamefont {Younis}}, \bibinfo {author}
  {\bibfnamefont {C.}~\bibnamefont {Iancu}}, \bibinfo {author} {\bibfnamefont
  {D.~I.}\ \bibnamefont {Santiago}}, \ and\ \bibinfo {author} {\bibfnamefont
  {I.}~\bibnamefont {Siddiqi}},\ }\href {\doibase
  10.1103/PhysRevResearch.4.033140} {\bibfield  {journal} {\bibinfo  {journal}
  {Phys. Rev. Res.}\ }\textbf {\bibinfo {volume} {4}},\ \bibinfo {pages}
  {033140} (\bibinfo {year} {2022})}\BibitemShut {NoStop}%
\bibitem [{\citenamefont {Jouzdani}\ \emph {et~al.}(2022)\citenamefont
  {Jouzdani}, \citenamefont {Johnson}, \citenamefont {Mucciolo},\ and\
  \citenamefont {Stetcu}}]{QITE_Jouzdani}%
  \BibitemOpen
  \bibfield  {author} {\bibinfo {author} {\bibfnamefont {P.}~\bibnamefont
  {Jouzdani}}, \bibinfo {author} {\bibfnamefont {C.~W.}\ \bibnamefont
  {Johnson}}, \bibinfo {author} {\bibfnamefont {E.~R.}\ \bibnamefont
  {Mucciolo}}, \ and\ \bibinfo {author} {\bibfnamefont {I.}~\bibnamefont
  {Stetcu}},\ }\href {\doibase 10.1103/PhysRevA.106.062435} {\bibfield
  {journal} {\bibinfo  {journal} {Phys. Rev. A}\ }\textbf {\bibinfo {volume}
  {106}},\ \bibinfo {pages} {062435} (\bibinfo {year} {2022})}\BibitemShut
  {NoStop}%
\bibitem [{\citenamefont {Choi}\ \emph {et~al.}(2021)\citenamefont {Choi},
  \citenamefont {Lee}, \citenamefont {Bonitati}, \citenamefont {Qian},\ and\
  \citenamefont {Watkins}}]{choi2021rodeo}%
  \BibitemOpen
  \bibfield  {author} {\bibinfo {author} {\bibfnamefont {K.}~\bibnamefont
  {Choi}}, \bibinfo {author} {\bibfnamefont {D.}~\bibnamefont {Lee}}, \bibinfo
  {author} {\bibfnamefont {J.}~\bibnamefont {Bonitati}}, \bibinfo {author}
  {\bibfnamefont {Z.}~\bibnamefont {Qian}}, \ and\ \bibinfo {author}
  {\bibfnamefont {J.}~\bibnamefont {Watkins}},\ }\href {\doibase
  10.1103/PhysRevLett.127.040505} {\bibfield  {journal} {\bibinfo  {journal}
  {Phys. Rev. Lett.}\ }\textbf {\bibinfo {volume} {127}},\ \bibinfo {pages}
  {040505} (\bibinfo {year} {2021})}\BibitemShut {NoStop}%
\bibitem [{\citenamefont {Coello~P\'erez}\ \emph {et~al.}(2022)\citenamefont
  {Coello~P\'erez}, \citenamefont {Bonitati}, \citenamefont {Lee},
  \citenamefont {Quaglioni},\ and\ \citenamefont {Wendt}}]{perez2022quantum}%
  \BibitemOpen
  \bibfield  {author} {\bibinfo {author} {\bibfnamefont {E.~A.}\ \bibnamefont
  {Coello~P\'erez}}, \bibinfo {author} {\bibfnamefont {J.}~\bibnamefont
  {Bonitati}}, \bibinfo {author} {\bibfnamefont {D.}~\bibnamefont {Lee}},
  \bibinfo {author} {\bibfnamefont {S.}~\bibnamefont {Quaglioni}}, \ and\
  \bibinfo {author} {\bibfnamefont {K.~A.}\ \bibnamefont {Wendt}},\ }\href
  {\doibase 10.1103/PhysRevA.105.032403} {\bibfield  {journal} {\bibinfo
  {journal} {Phys. Rev. A}\ }\textbf {\bibinfo {volume} {105}},\ \bibinfo
  {pages} {032403} (\bibinfo {year} {2022})}\BibitemShut {NoStop}%
\bibitem [{\citenamefont {Di~Matteo}\ \emph {et~al.}(2021)\citenamefont
  {Di~Matteo}, \citenamefont {McCoy}, \citenamefont {Gysbers}, \citenamefont
  {Miyagi}, \citenamefont {Woloshyn},\ and\ \citenamefont
  {Navr{\'a}til}}]{di2021improving}%
  \BibitemOpen
  \bibfield  {author} {\bibinfo {author} {\bibfnamefont {O.}~\bibnamefont
  {Di~Matteo}}, \bibinfo {author} {\bibfnamefont {A.}~\bibnamefont {McCoy}},
  \bibinfo {author} {\bibfnamefont {P.}~\bibnamefont {Gysbers}}, \bibinfo
  {author} {\bibfnamefont {T.}~\bibnamefont {Miyagi}}, \bibinfo {author}
  {\bibfnamefont {R.~M.}\ \bibnamefont {Woloshyn}}, \ and\ \bibinfo {author}
  {\bibfnamefont {P.}~\bibnamefont {Navr{\'a}til}},\ }\href {\doibase
  10.1103/PhysRevA.103.042405} {\bibfield  {journal} {\bibinfo  {journal}
  {Phys. Rev. A}\ }\textbf {\bibinfo {volume} {103}},\ \bibinfo {pages}
  {042405} (\bibinfo {year} {2021})}\BibitemShut {NoStop}%
\bibitem [{\citenamefont {Hlatshwayo}\ \emph {et~al.}(2022)\citenamefont
  {Hlatshwayo}, \citenamefont {Zhang}, \citenamefont {Wibowo}, \citenamefont
  {LaRose}, \citenamefont {Lacroix},\ and\ \citenamefont
  {Litvinova}}]{hlatshwayo2022simulating}%
  \BibitemOpen
  \bibfield  {author} {\bibinfo {author} {\bibfnamefont {M.~Q.}\ \bibnamefont
  {Hlatshwayo}}, \bibinfo {author} {\bibfnamefont {Y.}~\bibnamefont {Zhang}},
  \bibinfo {author} {\bibfnamefont {H.}~\bibnamefont {Wibowo}}, \bibinfo
  {author} {\bibfnamefont {R.}~\bibnamefont {LaRose}}, \bibinfo {author}
  {\bibfnamefont {D.}~\bibnamefont {Lacroix}}, \ and\ \bibinfo {author}
  {\bibfnamefont {E.}~\bibnamefont {Litvinova}},\ }\href {\doibase
  10.1103/PhysRevC.106.024319} {\bibfield  {journal} {\bibinfo  {journal}
  {Phys. Rev. C}\ }\textbf {\bibinfo {volume} {106}},\ \bibinfo {pages}
  {024319} (\bibinfo {year} {2022})}\BibitemShut {NoStop}%
\bibitem [{\citenamefont {Stetcu}\ \emph {et~al.}(2022)\citenamefont {Stetcu},
  \citenamefont {Baroni},\ and\ \citenamefont
  {Carlson}}]{stetcu2022variational}%
  \BibitemOpen
  \bibfield  {author} {\bibinfo {author} {\bibfnamefont {I.}~\bibnamefont
  {Stetcu}}, \bibinfo {author} {\bibfnamefont {A.}~\bibnamefont {Baroni}}, \
  and\ \bibinfo {author} {\bibfnamefont {J.}~\bibnamefont {Carlson}},\ }\href
  {\doibase 10.1103/PhysRevC.105.064308} {\bibfield  {journal} {\bibinfo
  {journal} {Phys. Rev. C}\ }\textbf {\bibinfo {volume} {105}},\ \bibinfo
  {pages} {064308} (\bibinfo {year} {2022})}\BibitemShut {NoStop}%
\bibitem [{\citenamefont {Dumitrescu}\ \emph {et~al.}(2018)\citenamefont
  {Dumitrescu}, \citenamefont {McCaskey}, \citenamefont {Hagen}, \citenamefont
  {Jansen}, \citenamefont {Morris}, \citenamefont {Papenbrock}, \citenamefont
  {Pooser}, \citenamefont {Dean},\ and\ \citenamefont
  {Lougovski}}]{dumitrescu2018cloud}%
  \BibitemOpen
  \bibfield  {author} {\bibinfo {author} {\bibfnamefont {E.~F.}\ \bibnamefont
  {Dumitrescu}}, \bibinfo {author} {\bibfnamefont {A.~J.}\ \bibnamefont
  {McCaskey}}, \bibinfo {author} {\bibfnamefont {G.}~\bibnamefont {Hagen}},
  \bibinfo {author} {\bibfnamefont {G.~R.}\ \bibnamefont {Jansen}}, \bibinfo
  {author} {\bibfnamefont {T.~D.}\ \bibnamefont {Morris}}, \bibinfo {author}
  {\bibfnamefont {T.}~\bibnamefont {Papenbrock}}, \bibinfo {author}
  {\bibfnamefont {R.~C.}\ \bibnamefont {Pooser}}, \bibinfo {author}
  {\bibfnamefont {D.~J.}\ \bibnamefont {Dean}}, \ and\ \bibinfo {author}
  {\bibfnamefont {P.}~\bibnamefont {Lougovski}},\ }\href {\doibase
  10.1103/PhysRevLett.120.210501} {\bibfield  {journal} {\bibinfo  {journal}
  {Phys. Rev. Lett.}\ }\textbf {\bibinfo {volume} {120}},\ \bibinfo {pages}
  {210501} (\bibinfo {year} {2018})}\BibitemShut {NoStop}%
\bibitem [{Ber()}]{Berkeley_AQT}%
  \BibitemOpen
  \href {https://aqt.lbl.gov} {\enquote {\bibinfo {title} {Berkeley advanced
  quantum testbed's website: https://aqt.lbl.gov},}\ }\BibitemShut {NoStop}%
\bibitem [{\citenamefont {Roggero}\ and\ \citenamefont
  {Carlson}(2019)}]{roggero2019dynamic}%
  \BibitemOpen
  \bibfield  {author} {\bibinfo {author} {\bibfnamefont {A.}~\bibnamefont
  {Roggero}}\ and\ \bibinfo {author} {\bibfnamefont {J.}~\bibnamefont
  {Carlson}},\ }\href {\doibase 10.1103/PhysRevC.100.034610} {\bibfield
  {journal} {\bibinfo  {journal} {Phys. Rev. C}\ }\textbf {\bibinfo {volume}
  {100}},\ \bibinfo {pages} {034610} (\bibinfo {year} {2019})}\BibitemShut
  {NoStop}%
\bibitem [{\citenamefont {Roggero}\ \emph {et~al.}(2020)\citenamefont
  {Roggero}, \citenamefont {Li}, \citenamefont {Carlson}, \citenamefont
  {Gupta},\ and\ \citenamefont {Perdue}}]{roggero2020quantum}%
  \BibitemOpen
  \bibfield  {author} {\bibinfo {author} {\bibfnamefont {A.}~\bibnamefont
  {Roggero}}, \bibinfo {author} {\bibfnamefont {A.~C.~Y.}\ \bibnamefont {Li}},
  \bibinfo {author} {\bibfnamefont {J.}~\bibnamefont {Carlson}}, \bibinfo
  {author} {\bibfnamefont {R.}~\bibnamefont {Gupta}}, \ and\ \bibinfo {author}
  {\bibfnamefont {G.~N.}\ \bibnamefont {Perdue}},\ }\href {\doibase
  10.1103/PhysRevD.101.074038} {\bibfield  {journal} {\bibinfo  {journal}
  {Phys. Rev. D}\ }\textbf {\bibinfo {volume} {101}},\ \bibinfo {pages}
  {074038} (\bibinfo {year} {2020})}\BibitemShut {NoStop}%
\bibitem [{\citenamefont {Baroni}\ \emph {et~al.}(2022)\citenamefont {Baroni},
  \citenamefont {Carlson}, \citenamefont {Gupta}, \citenamefont {Li},
  \citenamefont {Perdue},\ and\ \citenamefont {Roggero}}]{baroni2022nuclear}%
  \BibitemOpen
  \bibfield  {author} {\bibinfo {author} {\bibfnamefont {A.}~\bibnamefont
  {Baroni}}, \bibinfo {author} {\bibfnamefont {J.}~\bibnamefont {Carlson}},
  \bibinfo {author} {\bibfnamefont {R.}~\bibnamefont {Gupta}}, \bibinfo
  {author} {\bibfnamefont {A.~C.~Y.}\ \bibnamefont {Li}}, \bibinfo {author}
  {\bibfnamefont {G.~N.}\ \bibnamefont {Perdue}}, \ and\ \bibinfo {author}
  {\bibfnamefont {A.}~\bibnamefont {Roggero}},\ }\href {\doibase
  10.1103/PhysRevD.105.074503} {\bibfield  {journal} {\bibinfo  {journal}
  {Phys. Rev. D}\ }\textbf {\bibinfo {volume} {105}},\ \bibinfo {pages}
  {074503} (\bibinfo {year} {2022})}\BibitemShut {NoStop}%
\bibitem [{\citenamefont {Du}\ \emph {et~al.}(2021)\citenamefont {Du},
  \citenamefont {Vary}, \citenamefont {Zhao},\ and\ \citenamefont
  {Zuo}}]{inelastic_scattering}%
  \BibitemOpen
  \bibfield  {author} {\bibinfo {author} {\bibfnamefont {W.}~\bibnamefont
  {Du}}, \bibinfo {author} {\bibfnamefont {J.~P.}\ \bibnamefont {Vary}},
  \bibinfo {author} {\bibfnamefont {X.}~\bibnamefont {Zhao}}, \ and\ \bibinfo
  {author} {\bibfnamefont {W.}~\bibnamefont {Zuo}},\ }\href {\doibase
  10.1103/PhysRevA.104.012611} {\bibfield  {journal} {\bibinfo  {journal}
  {Phys. Rev. A}\ }\textbf {\bibinfo {volume} {104}},\ \bibinfo {pages}
  {012611} (\bibinfo {year} {2021})}\BibitemShut {NoStop}%
\bibitem [{\citenamefont {Bedaque}\ \emph {et~al.}(2022)\citenamefont
  {Bedaque}, \citenamefont {Khadka}, \citenamefont {Rupak},\ and\ \citenamefont
  {Yusf}}]{bedaque2022radiative}%
  \BibitemOpen
  \bibfield  {author} {\bibinfo {author} {\bibfnamefont {P.~F.}\ \bibnamefont
  {Bedaque}}, \bibinfo {author} {\bibfnamefont {R.}~\bibnamefont {Khadka}},
  \bibinfo {author} {\bibfnamefont {G.}~\bibnamefont {Rupak}}, \ and\ \bibinfo
  {author} {\bibfnamefont {M.}~\bibnamefont {Yusf}},\ }\href {\doibase
  https://doi.org/10.48550/arXiv.2209.09962} {\enquote {\bibinfo {title}
  {Radiative processes on a quantum computer},}\ } (\bibinfo {year} {2022}),\
  \Eprint {http://arxiv.org/abs/2209.09962} {arXiv:2209.09962 [nucl-th]}
  \BibitemShut {NoStop}%
\bibitem [{\citenamefont {Ciavarella}(2020)}]{Anthony_scattering}%
  \BibitemOpen
  \bibfield  {author} {\bibinfo {author} {\bibfnamefont {A.}~\bibnamefont
  {Ciavarella}},\ }\href {\doibase 10.1103/PhysRevD.102.094505} {\bibfield
  {journal} {\bibinfo  {journal} {Phys. Rev. D}\ }\textbf {\bibinfo {volume}
  {102}},\ \bibinfo {pages} {094505} (\bibinfo {year} {2020})}\BibitemShut
  {NoStop}%
\bibitem [{\citenamefont {Holland}\ \emph {et~al.}(2020)\citenamefont
  {Holland}, \citenamefont {Wendt}, \citenamefont {Kravvaris}, \citenamefont
  {Wu}, \citenamefont {Ormand}, \citenamefont {DuBois}, \citenamefont
  {Quaglioni},\ and\ \citenamefont {Pederiva}}]{Holland-2020}%
  \BibitemOpen
  \bibfield  {author} {\bibinfo {author} {\bibfnamefont {E.~T.}\ \bibnamefont
  {Holland}}, \bibinfo {author} {\bibfnamefont {K.~A.}\ \bibnamefont {Wendt}},
  \bibinfo {author} {\bibfnamefont {K.}~\bibnamefont {Kravvaris}}, \bibinfo
  {author} {\bibfnamefont {X.}~\bibnamefont {Wu}}, \bibinfo {author}
  {\bibfnamefont {W.~E.}\ \bibnamefont {Ormand}}, \bibinfo {author}
  {\bibfnamefont {J.~L.}\ \bibnamefont {DuBois}}, \bibinfo {author}
  {\bibfnamefont {S.}~\bibnamefont {Quaglioni}}, \ and\ \bibinfo {author}
  {\bibfnamefont {F.}~\bibnamefont {Pederiva}},\ }\href {\doibase
  10.1103/PhysRevA.101.062307} {\bibfield  {journal} {\bibinfo  {journal}
  {Phys. Rev. A}\ }\textbf {\bibinfo {volume} {101}},\ \bibinfo {pages}
  {062307} (\bibinfo {year} {2020})}\BibitemShut {NoStop}%
\bibitem [{\citenamefont {Hashim}\ \emph {et~al.}(2021)\citenamefont {Hashim},
  \citenamefont {Naik}, \citenamefont {Morvan}, \citenamefont {Ville},
  \citenamefont {Mitchell}, \citenamefont {Kreikebaum}, \citenamefont {Davis},
  \citenamefont {Smith}, \citenamefont {Iancu}, \citenamefont {O'Brien},
  \citenamefont {Hincks}, \citenamefont {Wallman}, \citenamefont {Emerson},\
  and\ \citenamefont {Siddiqi}}]{RC_Akel}%
  \BibitemOpen
  \bibfield  {author} {\bibinfo {author} {\bibfnamefont {A.}~\bibnamefont
  {Hashim}}, \bibinfo {author} {\bibfnamefont {R.~K.}\ \bibnamefont {Naik}},
  \bibinfo {author} {\bibfnamefont {A.}~\bibnamefont {Morvan}}, \bibinfo
  {author} {\bibfnamefont {J.-L.}\ \bibnamefont {Ville}}, \bibinfo {author}
  {\bibfnamefont {B.~K.}\ \bibnamefont {Mitchell}}, \bibinfo {author}
  {\bibfnamefont {J.~M.}\ \bibnamefont {Kreikebaum}}, \bibinfo {author}
  {\bibfnamefont {M.}~\bibnamefont {Davis}}, \bibinfo {author} {\bibfnamefont
  {E.}~\bibnamefont {Smith}}, \bibinfo {author} {\bibfnamefont
  {C.}~\bibnamefont {Iancu}}, \bibinfo {author} {\bibfnamefont {K.~P.}\
  \bibnamefont {O'Brien}}, \bibinfo {author} {\bibfnamefont {I.}~\bibnamefont
  {Hincks}}, \bibinfo {author} {\bibfnamefont {J.~J.}\ \bibnamefont {Wallman}},
  \bibinfo {author} {\bibfnamefont {J.}~\bibnamefont {Emerson}}, \ and\
  \bibinfo {author} {\bibfnamefont {I.}~\bibnamefont {Siddiqi}},\ }\href
  {\doibase 10.1103/PhysRevX.11.041039} {\bibfield  {journal} {\bibinfo
  {journal} {Phys. Rev. X}\ }\textbf {\bibinfo {volume} {11}},\ \bibinfo
  {pages} {041039} (\bibinfo {year} {2021})}\BibitemShut {NoStop}%
\bibitem [{\citenamefont {Feynman}\ and\ \citenamefont
  {Hibbs}(1965)}]{Feynman_saddlepoint}%
  \BibitemOpen
  \bibfield  {author} {\bibinfo {author} {\bibfnamefont {R.~P.}\ \bibnamefont
  {Feynman}}\ and\ \bibinfo {author} {\bibfnamefont {A.~R.}\ \bibnamefont
  {Hibbs}},\ }\href {https://cds.cern.ch/record/100771} {\emph {\bibinfo
  {title} {{Quantum mechanics and path integrals}}}},\ International series in
  pure and applied physics\ (\bibinfo  {publisher} {McGraw-Hill},\ \bibinfo
  {address} {New York, NY},\ \bibinfo {year} {1965})\BibitemShut {NoStop}%
\bibitem [{\citenamefont {Verlet}(1967)}]{verlet1967computer}%
  \BibitemOpen
  \bibfield  {author} {\bibinfo {author} {\bibfnamefont {L.}~\bibnamefont
  {Verlet}},\ }\href {\doibase 10.1103/PhysRev.159.98} {\bibfield  {journal}
  {\bibinfo  {journal} {Phys. Rev.}\ }\textbf {\bibinfo {volume} {159}},\
  \bibinfo {pages} {98} (\bibinfo {year} {1967})}\BibitemShut {NoStop}%
\bibitem [{\citenamefont {Swope}\ \emph {et~al.}(1982)\citenamefont {Swope},
  \citenamefont {Andersen}, \citenamefont {Berens},\ and\ \citenamefont
  {Wilson}}]{swope1982computer}%
  \BibitemOpen
  \bibfield  {author} {\bibinfo {author} {\bibfnamefont {W.~C.}\ \bibnamefont
  {Swope}}, \bibinfo {author} {\bibfnamefont {H.~C.}\ \bibnamefont {Andersen}},
  \bibinfo {author} {\bibfnamefont {P.~H.}\ \bibnamefont {Berens}}, \ and\
  \bibinfo {author} {\bibfnamefont {K.~R.}\ \bibnamefont {Wilson}},\ }\href
  {\doibase 10.1063/1.442716} {\bibfield  {journal} {\bibinfo  {journal} {J.
  Chem. Phys.}\ }\textbf {\bibinfo {volume} {76}},\ \bibinfo {pages} {637}
  (\bibinfo {year} {1982})}\BibitemShut {NoStop}%
\bibitem [{qis()}]{qiskit}%
  \BibitemOpen
  \href {https://qiskit.org/documentation} {\enquote {\bibinfo {title} {Qiskit:
  https://qiskit.org/documentation},}\ }\BibitemShut {NoStop}%
\bibitem [{\citenamefont {Beale}\ \emph {et~al.}(2020)\citenamefont {Beale},
  \citenamefont {Boone}, \citenamefont {Carignan-Dugas}, \citenamefont
  {Chytros}, \citenamefont {Dahlen}, \citenamefont {Dawkins}, \citenamefont
  {Emerson}, \citenamefont {Ferracin}, \citenamefont {Frey}, \citenamefont
  {Hincks}, \citenamefont {Hufnagel}, \citenamefont {Iyer}, \citenamefont
  {Jain}, \citenamefont {Kolbush}, \citenamefont {Ospadov}, \citenamefont
  {Pino}, \citenamefont {Qassim}, \citenamefont {Saunders}, \citenamefont
  {Skanes-Norman}, \citenamefont {Stasiuk}, \citenamefont {Wallman},
  \citenamefont {Winick},\ and\ \citenamefont {Wright}}]{True_Q}%
  \BibitemOpen
  \bibfield  {author} {\bibinfo {author} {\bibfnamefont {S.~J.}\ \bibnamefont
  {Beale}}, \bibinfo {author} {\bibfnamefont {K.}~\bibnamefont {Boone}},
  \bibinfo {author} {\bibfnamefont {A.}~\bibnamefont {Carignan-Dugas}},
  \bibinfo {author} {\bibfnamefont {A.}~\bibnamefont {Chytros}}, \bibinfo
  {author} {\bibfnamefont {D.}~\bibnamefont {Dahlen}}, \bibinfo {author}
  {\bibfnamefont {H.}~\bibnamefont {Dawkins}}, \bibinfo {author} {\bibfnamefont
  {J.}~\bibnamefont {Emerson}}, \bibinfo {author} {\bibfnamefont
  {S.}~\bibnamefont {Ferracin}}, \bibinfo {author} {\bibfnamefont
  {V.}~\bibnamefont {Frey}}, \bibinfo {author} {\bibfnamefont {I.}~\bibnamefont
  {Hincks}}, \bibinfo {author} {\bibfnamefont {D.}~\bibnamefont {Hufnagel}},
  \bibinfo {author} {\bibfnamefont {P.}~\bibnamefont {Iyer}}, \bibinfo {author}
  {\bibfnamefont {A.}~\bibnamefont {Jain}}, \bibinfo {author} {\bibfnamefont
  {J.}~\bibnamefont {Kolbush}}, \bibinfo {author} {\bibfnamefont
  {E.}~\bibnamefont {Ospadov}}, \bibinfo {author} {\bibfnamefont {J.~L.}\
  \bibnamefont {Pino}}, \bibinfo {author} {\bibfnamefont {H.}~\bibnamefont
  {Qassim}}, \bibinfo {author} {\bibfnamefont {J.}~\bibnamefont {Saunders}},
  \bibinfo {author} {\bibfnamefont {J.}~\bibnamefont {Skanes-Norman}}, \bibinfo
  {author} {\bibfnamefont {A.}~\bibnamefont {Stasiuk}}, \bibinfo {author}
  {\bibfnamefont {J.~J.}\ \bibnamefont {Wallman}}, \bibinfo {author}
  {\bibfnamefont {A.}~\bibnamefont {Winick}}, \ and\ \bibinfo {author}
  {\bibfnamefont {E.}~\bibnamefont {Wright}},\ }\href {\doibase
  https://doi.org/10.5281/zenodo.3945250} {\enquote {\bibinfo {title}
  {True-q},}\ } (\bibinfo {year} {2020})\BibitemShut {NoStop}%
\bibitem [{\citenamefont {Wallman}\ and\ \citenamefont
  {Emerson}(2016)}]{RC_original}%
  \BibitemOpen
  \bibfield  {author} {\bibinfo {author} {\bibfnamefont {J.~J.}\ \bibnamefont
  {Wallman}}\ and\ \bibinfo {author} {\bibfnamefont {J.}~\bibnamefont
  {Emerson}},\ }\href {\doibase 10.1103/PhysRevA.94.052325} {\bibfield
  {journal} {\bibinfo  {journal} {Phys. Rev. A}\ }\textbf {\bibinfo {volume}
  {94}},\ \bibinfo {pages} {052325} (\bibinfo {year} {2016})}\BibitemShut
  {NoStop}%
\bibitem [{\citenamefont {Vidal}\ and\ \citenamefont {Dawson}(2004)}]{Cartan}%
  \BibitemOpen
  \bibfield  {author} {\bibinfo {author} {\bibfnamefont {G.}~\bibnamefont
  {Vidal}}\ and\ \bibinfo {author} {\bibfnamefont {C.~M.}\ \bibnamefont
  {Dawson}},\ }\href {\doibase 10.1103/PhysRevA.69.010301} {\bibfield
  {journal} {\bibinfo  {journal} {Phys. Rev. A}\ }\textbf {\bibinfo {volume}
  {69}},\ \bibinfo {pages} {010301(R)} (\bibinfo {year} {2004})}\BibitemShut
  {NoStop}%
\bibitem [{\citenamefont {Nielsen}\ and\ \citenamefont
  {Chuang}(2011)}]{Nielsen_Chuang}%
  \BibitemOpen
  \bibfield  {author} {\bibinfo {author} {\bibfnamefont {M.~A.}\ \bibnamefont
  {Nielsen}}\ and\ \bibinfo {author} {\bibfnamefont {I.~L.}\ \bibnamefont
  {Chuang}},\ }\href@noop {} {\emph {\bibinfo {title} {Quantum Computation and
  Quantum Information: 10th Anniversary Edition}}},\ \bibinfo {edition} {10th}\
  ed.\ (\bibinfo  {publisher} {Cambridge University Press},\ \bibinfo {address}
  {USA},\ \bibinfo {year} {2011})\BibitemShut {NoStop}%
\bibitem [{\citenamefont {Otten}\ \emph {et~al.}(2019)\citenamefont {Otten},
  \citenamefont {Cortes},\ and\ \citenamefont
  {Gray}}]{otten2019noiseresilient}%
  \BibitemOpen
  \bibfield  {author} {\bibinfo {author} {\bibfnamefont {M.}~\bibnamefont
  {Otten}}, \bibinfo {author} {\bibfnamefont {C.~L.}\ \bibnamefont {Cortes}}, \
  and\ \bibinfo {author} {\bibfnamefont {S.~K.}\ \bibnamefont {Gray}},\
  }\href@noop {} {\enquote {\bibinfo {title} {Noise-resilient quantum dynamics
  using symmetry-preserving ansatzes},}\ } (\bibinfo {year} {2019})\BibitemShut
  {NoStop}%
\bibitem [{\citenamefont {Zhong}\ \emph {et~al.}(2020)\citenamefont {Zhong},
  \citenamefont {Wang}, \citenamefont {Deng}, \citenamefont {Chen},
  \citenamefont {Peng}, \citenamefont {Luo}, \citenamefont {Qin}, \citenamefont
  {Wu}, \citenamefont {Ding}, \citenamefont {Hu}, \citenamefont {Hu},
  \citenamefont {Yang}, \citenamefont {Zhang}, \citenamefont {Li},
  \citenamefont {Li}, \citenamefont {Jiang}, \citenamefont {Gan}, \citenamefont
  {Yang}, \citenamefont {You}, \citenamefont {Wang}, \citenamefont {Li},
  \citenamefont {Liu}, \citenamefont {Lu},\ and\ \citenamefont
  {Pan}}]{TVD_photons}%
  \BibitemOpen
  \bibfield  {author} {\bibinfo {author} {\bibfnamefont {H.-S.}\ \bibnamefont
  {Zhong}}, \bibinfo {author} {\bibfnamefont {H.}~\bibnamefont {Wang}},
  \bibinfo {author} {\bibfnamefont {Y.-H.}\ \bibnamefont {Deng}}, \bibinfo
  {author} {\bibfnamefont {M.-C.}\ \bibnamefont {Chen}}, \bibinfo {author}
  {\bibfnamefont {L.-C.}\ \bibnamefont {Peng}}, \bibinfo {author}
  {\bibfnamefont {Y.-H.}\ \bibnamefont {Luo}}, \bibinfo {author} {\bibfnamefont
  {J.}~\bibnamefont {Qin}}, \bibinfo {author} {\bibfnamefont {D.}~\bibnamefont
  {Wu}}, \bibinfo {author} {\bibfnamefont {X.}~\bibnamefont {Ding}}, \bibinfo
  {author} {\bibfnamefont {Y.}~\bibnamefont {Hu}}, \bibinfo {author}
  {\bibfnamefont {P.}~\bibnamefont {Hu}}, \bibinfo {author} {\bibfnamefont
  {X.-Y.}\ \bibnamefont {Yang}}, \bibinfo {author} {\bibfnamefont {W.-J.}\
  \bibnamefont {Zhang}}, \bibinfo {author} {\bibfnamefont {H.}~\bibnamefont
  {Li}}, \bibinfo {author} {\bibfnamefont {Y.}~\bibnamefont {Li}}, \bibinfo
  {author} {\bibfnamefont {X.}~\bibnamefont {Jiang}}, \bibinfo {author}
  {\bibfnamefont {L.}~\bibnamefont {Gan}}, \bibinfo {author} {\bibfnamefont
  {G.}~\bibnamefont {Yang}}, \bibinfo {author} {\bibfnamefont {L.}~\bibnamefont
  {You}}, \bibinfo {author} {\bibfnamefont {Z.}~\bibnamefont {Wang}}, \bibinfo
  {author} {\bibfnamefont {L.}~\bibnamefont {Li}}, \bibinfo {author}
  {\bibfnamefont {N.-L.}\ \bibnamefont {Liu}}, \bibinfo {author} {\bibfnamefont
  {C.-Y.}\ \bibnamefont {Lu}}, \ and\ \bibinfo {author} {\bibfnamefont {J.-W.}\
  \bibnamefont {Pan}},\ }\href {\doibase 10.1126/science.abe8770} {\bibfield
  {journal} {\bibinfo  {journal} {Science}\ }\textbf {\bibinfo {volume}
  {370}},\ \bibinfo {pages} {1460} (\bibinfo {year} {2020})}\BibitemShut
  {NoStop}%
\bibitem [{\citenamefont {Polyzou}\ and\ \citenamefont
  {Nathanson}(2020)}]{PhysRevC.101.064001}%
  \BibitemOpen
  \bibfield  {author} {\bibinfo {author} {\bibfnamefont {W.~N.}\ \bibnamefont
  {Polyzou}}\ and\ \bibinfo {author} {\bibfnamefont {E.}~\bibnamefont
  {Nathanson}},\ }\href {\doibase 10.1103/PhysRevC.101.064001} {\bibfield
  {journal} {\bibinfo  {journal} {Phys. Rev. C}\ }\textbf {\bibinfo {volume}
  {101}},\ \bibinfo {pages} {064001} (\bibinfo {year} {2020})}\BibitemShut
  {NoStop}%
\bibitem [{\citenamefont {Troyer}\ and\ \citenamefont
  {Wiese}(2005)}]{Troyer_fermion_sign}%
  \BibitemOpen
  \bibfield  {author} {\bibinfo {author} {\bibfnamefont {M.}~\bibnamefont
  {Troyer}}\ and\ \bibinfo {author} {\bibfnamefont {U.-J.}\ \bibnamefont
  {Wiese}},\ }\href {\doibase 10.1103/PhysRevLett.94.170201} {\bibfield
  {journal} {\bibinfo  {journal} {Phys. Rev. Lett.}\ }\textbf {\bibinfo
  {volume} {94}},\ \bibinfo {pages} {170201} (\bibinfo {year}
  {2005})}\BibitemShut {NoStop}%
\bibitem [{Iqu()}]{Iqus_website}%
  \BibitemOpen
  \href {https://iqus.uw.edu} {\enquote {\bibinfo {title}
  {https://iqus.uw.edu},}\ }\BibitemShut {NoStop}%
\bibitem [{\citenamefont {Huang}\ \emph
  {et~al.}(2020{\natexlab{b}})\citenamefont {Huang}, \citenamefont {Kueng},\
  and\ \citenamefont {Preskill}}]{huang2020predicting}%
  \BibitemOpen
  \bibfield  {author} {\bibinfo {author} {\bibfnamefont {H.-Y.}\ \bibnamefont
  {Huang}}, \bibinfo {author} {\bibfnamefont {R.}~\bibnamefont {Kueng}}, \ and\
  \bibinfo {author} {\bibfnamefont {J.}~\bibnamefont {Preskill}},\ }\href
  {\doibase 10.1038/s41567-020-0932-7} {\bibfield  {journal} {\bibinfo
  {journal} {Nat. Phys.}\ }\textbf {\bibinfo {volume} {16}},\ \bibinfo {pages}
  {1050} (\bibinfo {year} {2020}{\natexlab{b}})}\BibitemShut {NoStop}%
\bibitem [{\citenamefont {Koh}\ and\ \citenamefont
  {Grewal}(2022)}]{koh2022classical}%
  \BibitemOpen
  \bibfield  {author} {\bibinfo {author} {\bibfnamefont {D.~E.}\ \bibnamefont
  {Koh}}\ and\ \bibinfo {author} {\bibfnamefont {S.}~\bibnamefont {Grewal}},\
  }\href {\doibase 10.22331/q-2022-08-16-776} {\bibfield  {journal} {\bibinfo
  {journal} {Quantum}\ }\textbf {\bibinfo {volume} {6}},\ \bibinfo {pages}
  {776} (\bibinfo {year} {2022})}\BibitemShut {NoStop}%
\bibitem [{\citenamefont {Blais}\ \emph {et~al.}(2004)\citenamefont {Blais},
  \citenamefont {Huang}, \citenamefont {Wallraff}, \citenamefont {Girvin},\
  and\ \citenamefont {Schoelkopf}}]{cqed_Blais}%
  \BibitemOpen
  \bibfield  {author} {\bibinfo {author} {\bibfnamefont {A.}~\bibnamefont
  {Blais}}, \bibinfo {author} {\bibfnamefont {R.-S.}\ \bibnamefont {Huang}},
  \bibinfo {author} {\bibfnamefont {A.}~\bibnamefont {Wallraff}}, \bibinfo
  {author} {\bibfnamefont {S.~M.}\ \bibnamefont {Girvin}}, \ and\ \bibinfo
  {author} {\bibfnamefont {R.~J.}\ \bibnamefont {Schoelkopf}},\ }\href
  {\doibase 10.1103/PhysRevA.69.062320} {\bibfield  {journal} {\bibinfo
  {journal} {Phys. Rev. A}\ }\textbf {\bibinfo {volume} {69}},\ \bibinfo
  {pages} {062320} (\bibinfo {year} {2004})}\BibitemShut {NoStop}%
\bibitem [{\citenamefont {Wallraff}\ \emph {et~al.}(2004)\citenamefont
  {Wallraff}, \citenamefont {Schuster}, \citenamefont {Blais}, \citenamefont
  {Frunzio}, \citenamefont {Huang}, \citenamefont {Majer}, \citenamefont
  {Kumar}, \citenamefont {Girvin},\ and\ \citenamefont
  {Schoelkopf}}]{cqed_Wallraff}%
  \BibitemOpen
  \bibfield  {author} {\bibinfo {author} {\bibfnamefont {A.}~\bibnamefont
  {Wallraff}}, \bibinfo {author} {\bibfnamefont {D.~I.}\ \bibnamefont
  {Schuster}}, \bibinfo {author} {\bibfnamefont {A.}~\bibnamefont {Blais}},
  \bibinfo {author} {\bibfnamefont {L.}~\bibnamefont {Frunzio}}, \bibinfo
  {author} {\bibfnamefont {R.-S.}\ \bibnamefont {Huang}}, \bibinfo {author}
  {\bibfnamefont {J.}~\bibnamefont {Majer}}, \bibinfo {author} {\bibfnamefont
  {S.}~\bibnamefont {Kumar}}, \bibinfo {author} {\bibfnamefont {S.~M.}\
  \bibnamefont {Girvin}}, \ and\ \bibinfo {author} {\bibfnamefont {R.~J.}\
  \bibnamefont {Schoelkopf}},\ }\href {\doibase 10.1038/nature02851} {\bibfield
   {journal} {\bibinfo  {journal} {Nature}\ }\textbf {\bibinfo {volume}
  {431}},\ \bibinfo {pages} {162} (\bibinfo {year} {2004})}\BibitemShut
  {NoStop}%
\bibitem [{\citenamefont {Koch}\ \emph {et~al.}(2007)\citenamefont {Koch},
  \citenamefont {Yu}, \citenamefont {Gambetta}, \citenamefont {Houck},
  \citenamefont {Schuster}, \citenamefont {Majer}, \citenamefont {Blais},
  \citenamefont {Devoret}, \citenamefont {Girvin},\ and\ \citenamefont
  {Schoelkopf}}]{transmon}%
  \BibitemOpen
  \bibfield  {author} {\bibinfo {author} {\bibfnamefont {J.}~\bibnamefont
  {Koch}}, \bibinfo {author} {\bibfnamefont {T.~M.}\ \bibnamefont {Yu}},
  \bibinfo {author} {\bibfnamefont {J.}~\bibnamefont {Gambetta}}, \bibinfo
  {author} {\bibfnamefont {A.~A.}\ \bibnamefont {Houck}}, \bibinfo {author}
  {\bibfnamefont {D.~I.}\ \bibnamefont {Schuster}}, \bibinfo {author}
  {\bibfnamefont {J.}~\bibnamefont {Majer}}, \bibinfo {author} {\bibfnamefont
  {A.}~\bibnamefont {Blais}}, \bibinfo {author} {\bibfnamefont {M.~H.}\
  \bibnamefont {Devoret}}, \bibinfo {author} {\bibfnamefont {S.~M.}\
  \bibnamefont {Girvin}}, \ and\ \bibinfo {author} {\bibfnamefont {R.~J.}\
  \bibnamefont {Schoelkopf}},\ }\href {\doibase 10.1103/PhysRevA.76.042319}
  {\bibfield  {journal} {\bibinfo  {journal} {Phys. Rev. A}\ }\textbf {\bibinfo
  {volume} {76}},\ \bibinfo {pages} {042319} (\bibinfo {year}
  {2007})}\BibitemShut {NoStop}%
\bibitem [{\citenamefont {Bianchetti}\ \emph {et~al.}(2010)\citenamefont
  {Bianchetti}, \citenamefont {Filipp}, \citenamefont {Baur}, \citenamefont
  {Fink}, \citenamefont {Lang}, \citenamefont {Steffen}, \citenamefont
  {Boissonneault}, \citenamefont {Blais},\ and\ \citenamefont
  {Wallraff}}]{qutrit_Wallraff}%
  \BibitemOpen
  \bibfield  {author} {\bibinfo {author} {\bibfnamefont {R.}~\bibnamefont
  {Bianchetti}}, \bibinfo {author} {\bibfnamefont {S.}~\bibnamefont {Filipp}},
  \bibinfo {author} {\bibfnamefont {M.}~\bibnamefont {Baur}}, \bibinfo {author}
  {\bibfnamefont {J.~M.}\ \bibnamefont {Fink}}, \bibinfo {author}
  {\bibfnamefont {C.}~\bibnamefont {Lang}}, \bibinfo {author} {\bibfnamefont
  {L.}~\bibnamefont {Steffen}}, \bibinfo {author} {\bibfnamefont
  {M.}~\bibnamefont {Boissonneault}}, \bibinfo {author} {\bibfnamefont
  {A.}~\bibnamefont {Blais}}, \ and\ \bibinfo {author} {\bibfnamefont
  {A.}~\bibnamefont {Wallraff}},\ }\href {\doibase
  10.1103/PhysRevLett.105.223601} {\bibfield  {journal} {\bibinfo  {journal}
  {Phys. Rev. Lett.}\ }\textbf {\bibinfo {volume} {105}},\ \bibinfo {pages}
  {223601} (\bibinfo {year} {2010})}\BibitemShut {NoStop}%
\bibitem [{\citenamefont {Goss}\ \emph {et~al.}(2022)\citenamefont {Goss},
  \citenamefont {Morvan}, \citenamefont {Marinelli}, \citenamefont {Mitchell},
  \citenamefont {Nguyen}, \citenamefont {Naik}, \citenamefont {Chen},
  \citenamefont {Jünger}, \citenamefont {Kreikebaum}, \citenamefont
  {Santiago}, \citenamefont {Wallman},\ and\ \citenamefont
  {Siddiqi}}]{qutrit_Goss}%
  \BibitemOpen
  \bibfield  {author} {\bibinfo {author} {\bibfnamefont {N.}~\bibnamefont
  {Goss}}, \bibinfo {author} {\bibfnamefont {A.}~\bibnamefont {Morvan}},
  \bibinfo {author} {\bibfnamefont {B.}~\bibnamefont {Marinelli}}, \bibinfo
  {author} {\bibfnamefont {B.~K.}\ \bibnamefont {Mitchell}}, \bibinfo {author}
  {\bibfnamefont {L.~B.}\ \bibnamefont {Nguyen}}, \bibinfo {author}
  {\bibfnamefont {R.~K.}\ \bibnamefont {Naik}}, \bibinfo {author}
  {\bibfnamefont {L.}~\bibnamefont {Chen}}, \bibinfo {author} {\bibfnamefont
  {C.}~\bibnamefont {Jünger}}, \bibinfo {author} {\bibfnamefont {J.~M.}\
  \bibnamefont {Kreikebaum}}, \bibinfo {author} {\bibfnamefont {D.~I.}\
  \bibnamefont {Santiago}}, \bibinfo {author} {\bibfnamefont {J.~J.}\
  \bibnamefont {Wallman}}, \ and\ \bibinfo {author} {\bibfnamefont
  {I.}~\bibnamefont {Siddiqi}},\ }\href {\doibase 10.1038/s41467-022-34851-z}
  {\bibfield  {journal} {\bibinfo  {journal} {Nat. Commun.}\ }\textbf {\bibinfo
  {volume} {13}},\ \bibinfo {pages} {7481} (\bibinfo {year}
  {2022})}\BibitemShut {NoStop}%
\bibitem [{\citenamefont {Wallraff}\ \emph {et~al.}(2005)\citenamefont
  {Wallraff}, \citenamefont {Schuster}, \citenamefont {Blais}, \citenamefont
  {Frunzio}, \citenamefont {Majer}, \citenamefont {Devoret}, \citenamefont
  {Girvin},\ and\ \citenamefont
  {Schoelkopf}}]{single_dispersive_readout_Wallraff}%
  \BibitemOpen
  \bibfield  {author} {\bibinfo {author} {\bibfnamefont {A.}~\bibnamefont
  {Wallraff}}, \bibinfo {author} {\bibfnamefont {D.~I.}\ \bibnamefont
  {Schuster}}, \bibinfo {author} {\bibfnamefont {A.}~\bibnamefont {Blais}},
  \bibinfo {author} {\bibfnamefont {L.}~\bibnamefont {Frunzio}}, \bibinfo
  {author} {\bibfnamefont {J.}~\bibnamefont {Majer}}, \bibinfo {author}
  {\bibfnamefont {M.~H.}\ \bibnamefont {Devoret}}, \bibinfo {author}
  {\bibfnamefont {S.~M.}\ \bibnamefont {Girvin}}, \ and\ \bibinfo {author}
  {\bibfnamefont {R.~J.}\ \bibnamefont {Schoelkopf}},\ }\href {\doibase
  10.1103/PhysRevLett.95.060501} {\bibfield  {journal} {\bibinfo  {journal}
  {Phys. Rev. Lett.}\ }\textbf {\bibinfo {volume} {95}},\ \bibinfo {pages}
  {060501} (\bibinfo {year} {2005})}\BibitemShut {NoStop}%
\bibitem [{\citenamefont {Mallet}\ \emph {et~al.}(2009)\citenamefont {Mallet},
  \citenamefont {Ong}, \citenamefont {Palacios-Laloy}, \citenamefont {Nguyen},
  \citenamefont {Bertet}, \citenamefont {Vion},\ and\ \citenamefont
  {Esteve}}]{single_dispersive_readout_Mallet}%
  \BibitemOpen
  \bibfield  {author} {\bibinfo {author} {\bibfnamefont {F.}~\bibnamefont
  {Mallet}}, \bibinfo {author} {\bibfnamefont {F.~R.}\ \bibnamefont {Ong}},
  \bibinfo {author} {\bibfnamefont {A.}~\bibnamefont {Palacios-Laloy}},
  \bibinfo {author} {\bibfnamefont {F.}~\bibnamefont {Nguyen}}, \bibinfo
  {author} {\bibfnamefont {P.}~\bibnamefont {Bertet}}, \bibinfo {author}
  {\bibfnamefont {D.}~\bibnamefont {Vion}}, \ and\ \bibinfo {author}
  {\bibfnamefont {D.}~\bibnamefont {Esteve}},\ }\href {\doibase
  10.1038/nphys1400} {\bibfield  {journal} {\bibinfo  {journal} {Nat. Phys.}\
  }\textbf {\bibinfo {volume} {5}},\ \bibinfo {pages} {791} (\bibinfo {year}
  {2009})}\BibitemShut {NoStop}%
\bibitem [{\citenamefont {Walter}\ \emph {et~al.}(2017)\citenamefont {Walter},
  \citenamefont {Kurpiers}, \citenamefont {Gasparinetti}, \citenamefont
  {Magnard}, \citenamefont {Potočnik}, \citenamefont {Salathé}, \citenamefont
  {Pechal}, \citenamefont {Mondal}, \citenamefont {Oppliger}, \citenamefont
  {Eichler},\ and\ \citenamefont
  {Wallraff}}]{single_dispersive_readout_Walter}%
  \BibitemOpen
  \bibfield  {author} {\bibinfo {author} {\bibfnamefont {T.}~\bibnamefont
  {Walter}}, \bibinfo {author} {\bibfnamefont {P.}~\bibnamefont {Kurpiers}},
  \bibinfo {author} {\bibfnamefont {S.}~\bibnamefont {Gasparinetti}}, \bibinfo
  {author} {\bibfnamefont {P.}~\bibnamefont {Magnard}}, \bibinfo {author}
  {\bibfnamefont {A.}~\bibnamefont {Potočnik}}, \bibinfo {author}
  {\bibfnamefont {Y.}~\bibnamefont {Salathé}}, \bibinfo {author}
  {\bibfnamefont {M.}~\bibnamefont {Pechal}}, \bibinfo {author} {\bibfnamefont
  {M.}~\bibnamefont {Mondal}}, \bibinfo {author} {\bibfnamefont
  {M.}~\bibnamefont {Oppliger}}, \bibinfo {author} {\bibfnamefont
  {C.}~\bibnamefont {Eichler}}, \ and\ \bibinfo {author} {\bibfnamefont
  {A.}~\bibnamefont {Wallraff}},\ }\href {\doibase
  10.1103/PhysRevApplied.7.054020} {\bibfield  {journal} {\bibinfo  {journal}
  {Phys. Rev. Applied}\ }\textbf {\bibinfo {volume} {7}},\ \bibinfo {pages}
  {054020} (\bibinfo {year} {2017})}\BibitemShut {NoStop}%
\bibitem [{\citenamefont {Jeffrey}\ \emph {et~al.}(2014)\citenamefont
  {Jeffrey}, \citenamefont {Sank}, \citenamefont {Mutus}, \citenamefont
  {White}, \citenamefont {Kelly}, \citenamefont {Barends}, \citenamefont
  {Chen}, \citenamefont {Chen}, \citenamefont {Chiaro}, \citenamefont
  {Dunsworth}, \citenamefont {Megrant}, \citenamefont {O'Malley}, \citenamefont
  {Neill}, \citenamefont {Roushan}, \citenamefont {Vainsencher}, \citenamefont
  {Wenner}, \citenamefont {Cleland},\ and\ \citenamefont
  {Martinis}}]{Bandpass_purcell_exp}%
  \BibitemOpen
  \bibfield  {author} {\bibinfo {author} {\bibfnamefont {E.}~\bibnamefont
  {Jeffrey}}, \bibinfo {author} {\bibfnamefont {D.}~\bibnamefont {Sank}},
  \bibinfo {author} {\bibfnamefont {J.~Y.}\ \bibnamefont {Mutus}}, \bibinfo
  {author} {\bibfnamefont {T.~C.}\ \bibnamefont {White}}, \bibinfo {author}
  {\bibfnamefont {J.}~\bibnamefont {Kelly}}, \bibinfo {author} {\bibfnamefont
  {R.}~\bibnamefont {Barends}}, \bibinfo {author} {\bibfnamefont
  {Y.}~\bibnamefont {Chen}}, \bibinfo {author} {\bibfnamefont {Z.}~\bibnamefont
  {Chen}}, \bibinfo {author} {\bibfnamefont {B.}~\bibnamefont {Chiaro}},
  \bibinfo {author} {\bibfnamefont {A.}~\bibnamefont {Dunsworth}}, \bibinfo
  {author} {\bibfnamefont {A.}~\bibnamefont {Megrant}}, \bibinfo {author}
  {\bibfnamefont {P.~J.~J.}\ \bibnamefont {O'Malley}}, \bibinfo {author}
  {\bibfnamefont {C.}~\bibnamefont {Neill}}, \bibinfo {author} {\bibfnamefont
  {P.}~\bibnamefont {Roushan}}, \bibinfo {author} {\bibfnamefont
  {A.}~\bibnamefont {Vainsencher}}, \bibinfo {author} {\bibfnamefont
  {J.}~\bibnamefont {Wenner}}, \bibinfo {author} {\bibfnamefont {A.~N.}\
  \bibnamefont {Cleland}}, \ and\ \bibinfo {author} {\bibfnamefont {J.~M.}\
  \bibnamefont {Martinis}},\ }\href {\doibase 10.1103/PhysRevLett.112.190504}
  {\bibfield  {journal} {\bibinfo  {journal} {Phys. Rev. Lett.}\ }\textbf
  {\bibinfo {volume} {112}},\ \bibinfo {pages} {190504} (\bibinfo {year}
  {2014})}\BibitemShut {NoStop}%
\bibitem [{\citenamefont {Sete}\ \emph {et~al.}(2015)\citenamefont {Sete},
  \citenamefont {Martinis},\ and\ \citenamefont
  {Korotkov}}]{Bandpass_purcell_theory}%
  \BibitemOpen
  \bibfield  {author} {\bibinfo {author} {\bibfnamefont {E.~A.}\ \bibnamefont
  {Sete}}, \bibinfo {author} {\bibfnamefont {J.~M.}\ \bibnamefont {Martinis}},
  \ and\ \bibinfo {author} {\bibfnamefont {A.~N.}\ \bibnamefont {Korotkov}},\
  }\href {\doibase 10.1103/PhysRevA.92.012325} {\bibfield  {journal} {\bibinfo
  {journal} {Phys. Rev. A}\ }\textbf {\bibinfo {volume} {92}},\ \bibinfo
  {pages} {012325} (\bibinfo {year} {2015})}\BibitemShut {NoStop}%
\bibitem [{\citenamefont {Mitchell}\ \emph {et~al.}(2021)\citenamefont
  {Mitchell}, \citenamefont {Naik}, \citenamefont {Morvan}, \citenamefont
  {Hashim}, \citenamefont {Kreikebaum}, \citenamefont {Marinelli},
  \citenamefont {Lavrijsen}, \citenamefont {Nowrouzi}, \citenamefont
  {Santiago},\ and\ \citenamefont {Siddiqi}}]{diff-AC-Stark}%
  \BibitemOpen
  \bibfield  {author} {\bibinfo {author} {\bibfnamefont {B.~K.}\ \bibnamefont
  {Mitchell}}, \bibinfo {author} {\bibfnamefont {R.~K.}\ \bibnamefont {Naik}},
  \bibinfo {author} {\bibfnamefont {A.}~\bibnamefont {Morvan}}, \bibinfo
  {author} {\bibfnamefont {A.}~\bibnamefont {Hashim}}, \bibinfo {author}
  {\bibfnamefont {J.~M.}\ \bibnamefont {Kreikebaum}}, \bibinfo {author}
  {\bibfnamefont {B.}~\bibnamefont {Marinelli}}, \bibinfo {author}
  {\bibfnamefont {W.}~\bibnamefont {Lavrijsen}}, \bibinfo {author}
  {\bibfnamefont {K.}~\bibnamefont {Nowrouzi}}, \bibinfo {author}
  {\bibfnamefont {D.~I.}\ \bibnamefont {Santiago}}, \ and\ \bibinfo {author}
  {\bibfnamefont {I.}~\bibnamefont {Siddiqi}},\ }\href {\doibase
  10.1103/PhysRevLett.127.200502} {\bibfield  {journal} {\bibinfo  {journal}
  {Phys. Rev. Lett.}\ }\textbf {\bibinfo {volume} {127}},\ \bibinfo {pages}
  {200502} (\bibinfo {year} {2021})}\BibitemShut {NoStop}%
\bibitem [{\citenamefont {Erhard}\ \emph {et~al.}(2019)\citenamefont {Erhard},
  \citenamefont {Wallman}, \citenamefont {Postler}, \citenamefont {Meth},
  \citenamefont {Stricker}, \citenamefont {Martinez}, \citenamefont
  {Schindler}, \citenamefont {Monz}, \citenamefont {Emerson},\ and\
  \citenamefont {Blatt}}]{cycle_benchmarking}%
  \BibitemOpen
  \bibfield  {author} {\bibinfo {author} {\bibfnamefont {A.}~\bibnamefont
  {Erhard}}, \bibinfo {author} {\bibfnamefont {J.~J.}\ \bibnamefont {Wallman}},
  \bibinfo {author} {\bibfnamefont {L.}~\bibnamefont {Postler}}, \bibinfo
  {author} {\bibfnamefont {M.}~\bibnamefont {Meth}}, \bibinfo {author}
  {\bibfnamefont {R.}~\bibnamefont {Stricker}}, \bibinfo {author}
  {\bibfnamefont {E.~A.}\ \bibnamefont {Martinez}}, \bibinfo {author}
  {\bibfnamefont {P.}~\bibnamefont {Schindler}}, \bibinfo {author}
  {\bibfnamefont {T.}~\bibnamefont {Monz}}, \bibinfo {author} {\bibfnamefont
  {J.}~\bibnamefont {Emerson}}, \ and\ \bibinfo {author} {\bibfnamefont
  {R.}~\bibnamefont {Blatt}},\ }\href {\doibase 10.1038/s41467-019-13068-7}
  {\bibfield  {journal} {\bibinfo  {journal} {Nat. Commun.}\ }\textbf {\bibinfo
  {volume} {10}},\ \bibinfo {pages} {5347} (\bibinfo {year}
  {2019})}\BibitemShut {NoStop}%
\end{thebibliography}%
\end{document}